\documentclass[sigconf,nonacm, natbib=true,anonymous=False]{acmart}
\setcopyright{none}
\usepackage{bm}
\usepackage{multirow}
\usepackage{enumitem}
\usepackage[normalem]{ulem}
\usepackage{subcaption}

\usepackage{etoc}
\etocdepthtag.toc{mtchapter}
\etocsettagdepth{mtchapter}{subsection}
\etocsettagdepth{mtappendix}{none}
\AtBeginDocument{%
  }

\acmISBN{978-1-4503-XXXX-X/2018/06}

\usepackage{algorithm}
\usepackage{algpseudocode}
\begin{document}


\title{Evolutionary Router Feature Generation for Zero-Shot Graph Anomaly Detection with Mixture-of-Experts}

\author{Haiyang Jiang}
\affiliation{%
  \institution{The University of Queensland}
  \city{Brisbane}
  \state{Queensland}
  \country{Australia}
}
\email{haiyang.jiang@uq.edu.au}

\author{Tong Chen}
\affiliation{%
  \institution{The University of Queensland}
  \city{Brisbane}
  \state{Queensland}
  \country{Australia}}
\email{tong.chen@uq.edu.au}

\author{Xinyi Gao}
\affiliation{%
  \institution{The University of Queensland}
  \city{Brisbane}
  \state{Queensland}
  \country{Australia}}
\email{xinyi.gao@uq.edu.au}

\author{Guansong Pang}
\affiliation{%
  \institution{Singapore Management University}
  \country{Singapore}}
\email{gspang@smu.edu}

\author{Quoc Viet Hung Nguyen}
\affiliation{%
  \institution{The University of Queensland}
  \city{Gold Coast}
  \state{Queensland}
  \country{Australia}}
\email{henry.nguyen@griffith.edu.au}

\author{Hongzhi Yin}
\affiliation{%
  \institution{The University of Queensland}
  \city{Brisbane}
  \state{Queensland}
  \country{Australia}}
\email{db.hongzhi@gmail.com }
\renewcommand{\shortauthors}{Trovato et al.}

\begin{abstract}

Zero-shot graph anomaly detection (GAD) has attracted increasing attention recent years, yet the heterogeneity of graph structures, features, and anomaly patterns across graphs make existing single GNN methods insufficiently expressive to model diverse anomaly mechanisms. In this regard, Mixture-of-experts (MoE) architectures provide a promising paradigm by integrating diverse GNN experts with complementary inductive biases, yet their effectiveness in zero-shot GAD is severely constrained by distribution shifts, leading to two key routing challenges.
First, nodes often carry vastly different semantics across graphs, and straightforwardly performing routing based on their features is prone to generating biased or suboptimal expert assignments.
Second, as anomalous graphs often exhibit pronounced distributional discrepancies, existing router designs fall short in capturing domain-invariant routing principles that generalize beyond the training graphs. To address these challenges, we propose a novel MoE framework with \underline{evo}lutionary router \underline{f}eature \underline{g}eneration (EvoFG) for zero-shot GAD. To enhance MoE routing, we propose an evolutionary feature generation scheme that iteratively constructs and selects informative structural features via an LLM-based generator and Shapley-guided evaluation. Moreover, a memory-enhanced router with an invariant learning objective is designed to capture transferable routing patterns under distribution shifts. Extensive experiments on six benchmarks show that EvoFG consistently outperforms state-of-the-art baselines, achieving strong and stable zero-shot GAD performance.

\end{abstract}

\maketitle

\section{Introduction}
Graph anomaly detection (GAD) aims to identify anomalous nodes in a graph whose attributes or structural behaviors deviate significantly from the majority.
Such anomalous nodes often correspond to rare or unexpected events, including fraudulent users in financial networks, malfunctioning sensors in cyber-physical systems, or malicious accounts in social platforms~\cite{akoglu2015graph,ma2021comprehensive,pang2021deep,qiao2025deep,wang2026pamas}.
Accurate anomaly detection is therefore critical for a wide range of real-world applications.
In the conventional setting, most GAD methods are designed under a dataset-specific paradigm, where models are trained and evaluated on the same graph, relying on domain-dependent patterns to distinguish abnormal nodes from normal ones~\cite{ding2019deep,huang2022hop,liu2021anomaly}.

In many realistic scenarios, anomaly detection systems are expected to operate on new or unseen graphs, where retraining is impractical due to privacy constraints, limited annotations, or computational costs.
This has motivated growing interest in generalist GAD, which seeks to build a single framework capable of generalizing across multiple graph domains in zero-shot settings without retraining or finetuning. 
For example, \cite{niu2024zero} proposes a pretrained Graph Neural Network (GNN) that leverages neighborhood node prompts to capture transferable structural patterns across graphs. Then, a residual graph encoder is proposed in ARC \cite{liu2024arc} to capture how each node deviates from its neighbors, followed by an anomaly scoring function that measures the distance between each node's embedding and the ones reconstructed from globally sampled normal nodes. Recently, AnomalyGFM \cite{qiao2025anomalygfm} introduces an enhancement by pre-training a GNNConcepts to learn universal prototypes of normal and abnormal nodes, which are used as references during zero-shot prediction. 

Notably, all the aforementioned zero-shot GAD methods rely on a universal GNN model to extract node representations. 
However, in zero-shot GAD where a variety of graphs are handled during inference, sticking to a single GNN architecture is inherently biased toward a specific structural assumption and inductive bias, and is therefore often insufficient to capture the diverse and heterogeneous manifestations of abnormal nodes across different graph domains. For instance, the GNNs in both ARC and AnomalyGFM compute a residual term between the embeddings of the target node and its neighboring nodes, which operate under the homophily assumption. That is, a graph's local structures consist of similar nodes, thus abnormal nodes are likely to exhibit highly distinct patterns compared with their neighbors. As a result, GNNs designed with such an assumption can hardly generalize to graphs that exhibit heterophily (e.g., social networks with diverse users and cross-disciplinary citation networks) rather than homophily \cite{he2021bernnet,song2023ordered}. Such structural differences are also observed within the same graph. For example, aggregating higher-order neighbor information helps when embedding a node with sparse neighbors \cite{yun2019graph}, while a target node situated in a dense local structure may benefit from attentive message passing \cite{velivckovic2018graph} to counteract the noise from irrelevant neighbors. 
Hence, resorting to a single GNN for zero-shot GAD impedes model expressiveness when capturing complex structural or semantic abnormalities, defeating the purpose of a generalist framework.

In this regard, a natural strategy to enhance model expressiveness for zero-shot GAD is to integrate multiple GNNs that specialize in capturing complementary graph characteristics (e.g., homophily/heterophily, long-/short-range dependencies, etc.). To facilitate this, Mixture-of-Experts (MoE) \cite{wang2023graph} offers a principled framework. In an MoE framework containing diverse GNN experts, once a target node with its local structure is received, a router adaptively predicts the fitness of all experts, and the optimal ensemble of expert-generated embeddings is computed via either discrete selection or weighted aggregation. 
In MoE, generalization performance depends on two key factors: the quality of the experts~\cite{lee2024testam} and the ability of the router to identify appropriate experts for each target node~\cite{wang2023graph, guo2024dynamic,huang2025graph}.
While high-quality experts can typically be obtained through a set of carefully chosen GNNs, learning a reliable router remains challenging, especially in the zero-shot GAD setting. 
On the one hand, to make accurate decisions, the MoE router expects informative features describing the characteristics of each target node, which we term \textit{router features}. This renders raw node attributes or GNN-generated node embeddings an ineffective choice for router features, as the former is non-generalizable across graphs and the latter makes the router features inherently biased toward a specific GNN architecture. On the other hand, as routing decisions must be made for nodes from unseen graphs without domain-specific supervision, the router needs to learn causal, graph-invariant signals from the input router features. However, this capability is largely overlooked as existing MoE methods mostly perform training and inference on a single graph \cite{wang2023graph,hu2022graph,ma2024mixture}, which renders the router prone to misrouting target nodes from unseen graphs to inappropriate experts, eventually leading to subpar node embeddings and erroneous anomaly predictions in zero-shot GAD.

To address this bottleneck, we propose a novel MoE framework with \underline{evo}lutionary router \underline{f}eature \underline{g}eneration (EvoFG), designed to robustly route unseen node samples to appropriate experts in zero-shot GAD. 
In order to discover quality router features, we propose an automated feature engineering paradigm that iteratively generates and refines router features. Specifically, for each target node, we define a set of standardized features that characterize it w.r.t. both local and global graph structures (e.g., similarity with local neighbors and graph-level PageRank \cite{page1999pagerank} score), based on which a large language model (LLM) is utilized to generate new features via a chain-of-thought process. To ensure utility of newly generated features for expert routing, we perform feature selection by measuring the importance of each candidate router feature. This is achieved through the lens of a cooperative game theory, namely Shapley value \cite{chen2023algorithms} that computes each feature's marginal contribution by examining a model's performance under all possible feature combinations. To bypass the prohibitive cost of Shapley value computation in our MoE framework, we further propose a fast yet accurate approximation, where a sampling approach is in use to avoid enumerating all feature permutations, and a lightweight proxy task is designed to substitute the full MoE training cycle and quickly evaluate the router's performance with a given feature set. 

At the same time, in EvoFG, we revamp the inner mechanisms of the router, enabling it to learn the most critical routing signals from the router features and maintain robustness in zero-shot GAD. Notably, we introduce a memory-enhanced router architecture, allowing crucial routing-related information accumulated during the feature evolution process to be dynamically stored and retrieved in its memory banks during the router feature generation process. 
In addition, to prevent the router from picking up spurious correlations from generated router features, we put forward an innovative invariant learning paradigm specifically designed for the MoE router. With augmented environments created for each target node by partially masking the router features, we formulate an invariant risk minimization objective to encourage the router to learn causal, and ideally graph-invariant routing principles.
Paired with the automated router feature generation scheme, EvoFG continuously refines its routing efficacy, thereby substantially improving the anomaly detection accuracy on unseen graphs.

To summarize, the contributions of this work are three-fold:
\begin{itemize}[leftmargin=*]
    \item We reveal a fundamental bottleneck in existing generalist graph anomaly detection (GAD) models: the limited generalizability of the router caused by an insufficient and static router feature space, which prevents the router from learning transferable routing patterns across heterogeneous graph domains.
    
    \item We propose \textbf{EvoFG}, an iterative feature evolution framework that progressively generates and selects informative router features, integrates a memory-enhanced router to accumulate critical routing knowledge across iterations, and employs invariant learning to improve router generalization across feature-subset environments.
    \item Extensive experiments on six public benchmark datasets demonstrate that EvoFG consistently outperforms state-of-the-art generalist GAD methods.
\end{itemize}

\section{Preliminaries}
In this section, we first revisit the fundamental concept of  graph anomaly detection (GAD), and then formally define the problem we studied: zero-shot generalist GAD.
Due to space constraints, we place our comprehensive review on related work in \textbf{Appendix~\ref{RW}}.

\subsection{Graph Anomaly Detection}

We consider an attributed graph $\mathcal{G}=(\mathcal{V},\mathcal{E})$, where $\mathcal{V}=\{v_1,\dots,v_N\}$ denotes the node set and 
$\mathcal{E}\subseteq \mathcal{V}\times\mathcal{V}$ denotes the edge set. 
Node attributes are represented by a feature matrix $\mathbf{X}$, and the graph structure is encoded by an adjacency matrix $\mathbf{A}\in\{0,1\}^{N\times N}$.
The nodes are partitioned into a normal set $\mathcal{V}_n$ and an anomaly set $\mathcal{V}_a$, such that $\mathcal{V}_n \cup \mathcal{V}_a=\mathcal{V}$ and $\mathcal{V}_n \cap \mathcal{V}_a=\emptyset$, with $|\mathcal{V}_a|\ll |\mathcal{V}_n|$. An anomaly label vector $\mathbf{y}\in\{0,1\}^N$ is defined for all $N$ nodes, where $y_i=1$ if $v_i\in\mathcal{V}_a$ and $y_i=0$ otherwise.

The traditional graph anomaly detection (GAD) aims to learn a scoring function $f:\mathcal{V}\rightarrow\mathbb{R}$ that assigns higher scores to anomalous nodes than to normal ones, i.e., $f(v_a)>f(v_n)$ for $v_a\in\mathcal{V}_a$ and $v_n\in\mathcal{V}_n$.
The conventional GAD methods adopt a single-graph setting, where the model is trained and evaluated on the same dataset $\mathcal{D}=\{ \mathcal{G},\mathbf{y\}}$, assuming consistent anomaly patterns between training and inference.

\subsection{Problem Formulation}
In this paper, we focus on a more challenging and practical problem: \textbf{zero-shot  generalist GAD}, which aims to learn an anomaly scoring function $f$ with labeled source graphs that generalizes \textbf{across multiple target graphs from diverse domains}.
Let the source labeled collection be $\mathcal{T}_{s}=\{\mathcal{D}_i\}_{i=1}^{T}$, and the target collection be $\mathcal{T}_{t}=\{\mathcal{D}_j\}_{j=1}^{T'}$, 
where $\mathcal{T}_{s}\cap\mathcal{T}_{t}=\emptyset$ and datasets are sampled from different domains.
The scoring function $f$ is trained solely on the source collection $\mathcal{T}_{s}$ and directly applied to all target datasets in $\mathcal{T}_{t}$.
No anomaly labels from the target collection $\mathcal{T}_{t}$ are accessible at both training and inference stages.

\section{Methodology}

\begin{figure*}[t]
\setlength{\abovecaptionskip}{1pt}

    \centering
    \includegraphics[width=0.88\linewidth]{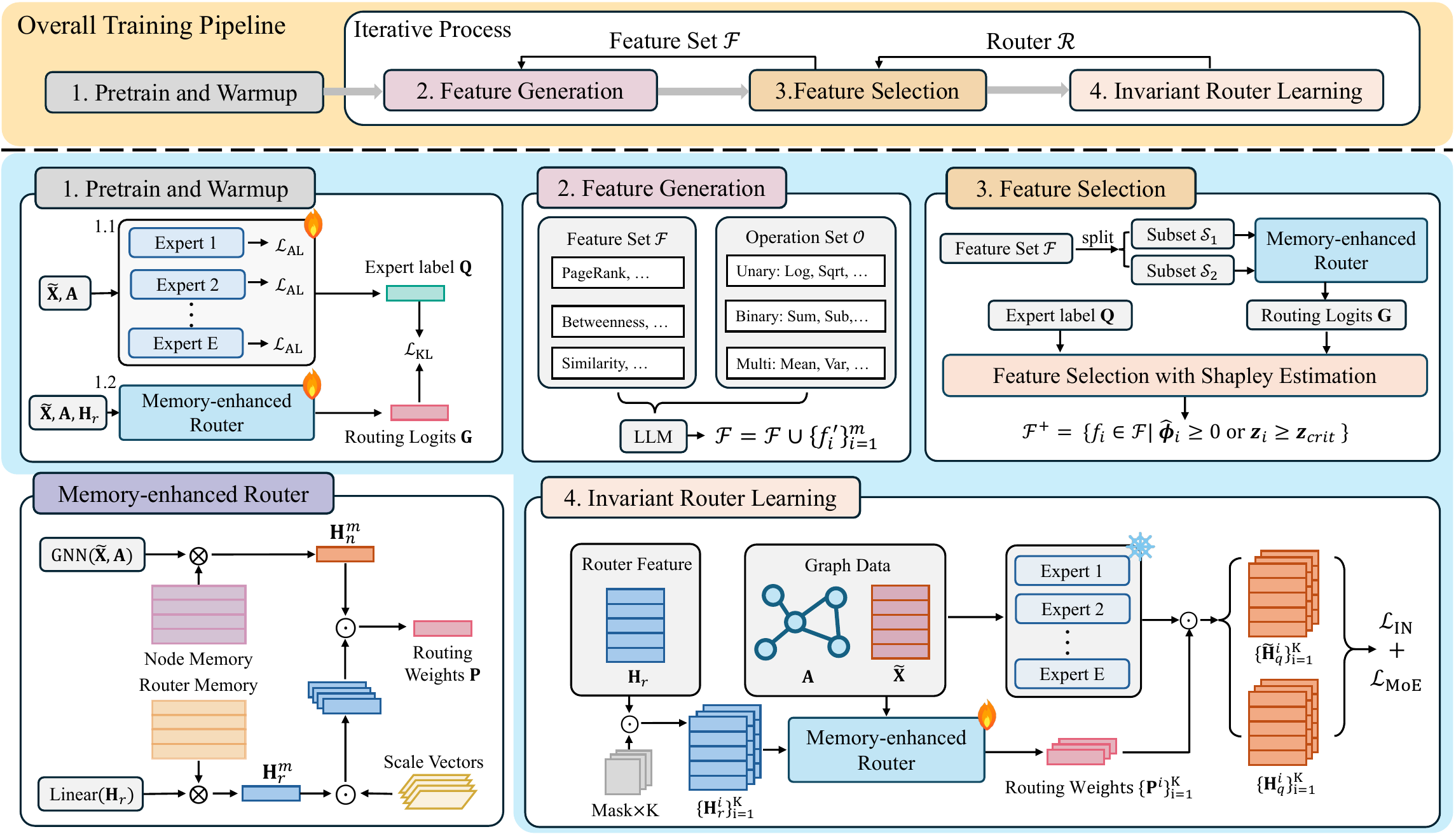}
    \caption{
Overall framework of EvoFG. Top: the iterative training pipeline. Bottom: detailed workflow.
}
    \label{fig:framework}
\end{figure*}

We hereby present our proposed EvoFG, which consists of automated router feature generation and a memory enhanced mixture of experts (MoE) architecture, as illustrated in Figure~\ref{fig:framework}. EvoFG leverages an LLM to generate novel router features from standardized primitives and applies accelerated Shapley estimation for feature evaluation and selection. Meanwhile, a memory enhanced router is introduced to retain critical routing information. The framework is optimized with an invariant learning objective to enable robust, environment independent routing across diverse graphs.

\subsection{GNN Mixture-of-Experts}

We employ a mixture-of-experts (MoE) framework comprising multiple GNN experts with diverse architectures to capture heterogeneous structural and semantic patterns across graph domains.

\noindent\textbf{Attribute Projection.}
Since node attributes across datasets encode different semantics, GNN models suffer from inconsistent feature dimensionality. To address this issue, we follow previous studies~\cite{liu2024arc,qiao2025anomalygfm,niu2024zero} to align node features. Specifically, we first project node attributes from different datasets into a unified latent dimension using PCA, denoted as: $\hat{\mathbf{X}}=\text{PCA}(\mathbf{X})$, where $\mathbf{X}\in\mathbb{R}^{N\times d_{ori}}$ and $\mathbf{\hat X}\in\mathbb{R}^{N\times d}$ with $d<d_{ori}$.
Despite dimensional alignment, attributes across graphs still encode inconsistent semantics, as principal components with identical rankings can correspond to different meanings.
Thus, the attributes are resorted based on the smoothness score \cite{liu2024arc}:
\begin{equation}
s_k(\hat{\mathbf{X}})
= -\frac{1}{|\mathcal{E}|}
\sum_{(v_i, v_j) \in \mathcal{E}}
\left( \mathbf{\hat{X}}_{i,k} - \mathbf{\hat{X}}_{j,k} \right)^2,
\end{equation}
where lower $s_k$ values indicate larger variations along graph edges, corresponding to high-frequency signals and stronger heterophily. Then, we align attribute dimensions by sorting attributes in an ascending order of smoothness scores, and we denote the sorted features as $\tilde{\mathbf{X}}\in\mathbb{R}^{N\times d}$. As per \cite{liu2024arc,qiao2025anomalygfm}, $\tilde{\mathbf{X}}$ enforces consistent structural frequency patterns across domains and thereby mitigates semantic misalignment.

\noindent\textbf{Cross-attention Encoding.}
Afterwards, each node is encoded by multiple GNN experts with diverse architectures:
\begin{equation}
\mathbf{H}^{e} = \mathrm{GNN}_e(\tilde{\mathbf{X}}, \mathbf{A}),
\end{equation}
where \(\mathbf{H}^{e}\in \mathbb{R}^{N\times d_e}\) indicates the node embeddings produced by the $e$-th expert. 
We incorporate multiple GNN experts with distinct graph convolution mechanisms, such as low pass filtering, high pass filtering, and attention based operators, to capture diverse anomaly characteristics and yield complementary structural information. The expert configurations are described in \textbf{Appendix~\ref{appendix:experts_intro}}.

To achieve zero-shot prediction, we enhance each GNN expert by an in-context cross-attention module that learns to distinguish anomalous nodes from normal ones. 
During training, a subset of normal nodes is randomly sampled as in-context samples, while the remaining nodes serve as query nodes. The query nodes' embeddings are reconstructed via cross-attention. Specifically, the expert representations are partitioned as $\mathbf{H}^{e} = \big[\mathbf{H}^{e}_k;\, \mathbf{H}^{e}_q\big]$, 
where \(\mathbf{H}^{e}_k\in \mathbb{R}^{N_k \times d}\) and \(\mathbf{H}^{e}_q\in \mathbb{R}^{N_q \times d}\) denote the in-context and query embeddings, respectively.
Then, in each expert, the query embeddings are reconstructed from the embeddings of in-context samples:
\begin{equation}
\label{eq:single_head_attention}
\begin{aligned}
&\mathbf{Q}^{e} = \mathbf{H}^{e}_q \mathbf{W}^{e}_q, \ \ \ 
\mathbf{K}^{e} = \mathbf{H}^{e}_k \mathbf{W}^{e}_k, \\
&\tilde{\mathbf{H}}^{e}_q
=
\mathrm{softmax}\!\left(
\frac{\mathbf{Q}^{e} \mathbf{K}^{e\top}}{\sqrt{d}}
\right)\mathbf{H}^{e}_k,
\end{aligned}
\end{equation}
where \(\mathbf{Q}^{e}\in \mathbb{R}^{N_q\times d'}\) and \(\mathbf{K}^{e}\in \mathbb{R}^{N_k\times d'}\) are query and key matrices. \(\mathbf{W}^{e}_q\) and \(\mathbf{W}^{e}_k \in \mathbb{R}^{d \times d'}\) are trainable parameters in each expert. Notably, we do not introduce the value matrix commonly used in other cross-attention \cite{vaswani2017attention,rombach2022high}, maintaining the reconstructed embeddings $\tilde{\mathbf{H}}^{e}_q$ in the same space as $\mathbf{H}^{e}_q$.

\noindent\textbf{Anomaly Detection via MoE.}
Afterwards, we propose to aggregate embeddings $\mathbf{H}_q^{e}$ and $\tilde{\mathbf{H}}_q^{e}$ via a soft aggregation routing strategy:
\begin{equation}
\label{eq:expert_aggregation}
\mathbf{H}_q^{\mathrm{final}}=\sum_{e=1}^{E} \mathbf{P}_{:,e}\mathbf{H}_q^{e}, \quad
\tilde{\mathbf{H}}_q^{\mathrm{final}}=\sum_{e=1}^{E} \mathbf{P}_{:,e}\tilde{\mathbf{H}}_q^{e},
\end{equation}
where $\mathbf{P}=\mathrm{Softmax}(\mathbf{G})$ represents the routing weight matrix obtained from the routing logits $\mathbf{G}\in\mathbb{R}^{N\times E}$, which will be detailed later in Section \ref{sec:mem}.



Unlike prior MoE models that rely on discrete expert selection \cite{shazeer2017outrageously,wang2023graph}, we employ a soft routing mechanism that aggregates all expert outputs. Discrete routing is inherently noise-sensitive and prone to biased assignments, which is particularly unstable in zero-shot GAD. In contrast, soft routing dynamically reweights experts to exploit complementary inductive biases and improve robustness under distribution shift, while introducing negligible overhead due to the small number of parallel experts (4 in our case). 

Finally, the anomaly scores for all query nodes are computed as the reconstruction discrepancy:
\begin{equation}
\label{eq:anomaly_score}
{c}=\left\|\mathbf{H}_q^{\mathrm{final}}-\tilde{\mathbf{H}}_q^{\mathrm{final}}\right\|_2,
\end{equation}
where $\left\|  \cdot  \right\|$ denotes the $\ell_2$ norm.
Intuitively, normal nodes exhibit small reconstruction errors due to shared latent patterns with in-context embeddings, whereas anomalous nodes deviate substantially, leading to larger scores.




\noindent\textbf{Decoupled Training of MoE.}
Instead of end to end MoE training, we adopt a two-stage optimization strategy, where experts are pretrained independently and the router is trained with frozen experts. This design mitigates routing collapse toward a dominant expert, improves training stability, and preserves expert diversity, consistent with our empirical results. 

To train the experts for anomaly detection, we adopt a cosine consistency loss to regulate the similarity between the GNN embeddings $\mathbf{H}_{q}^{e}$ and its reconstructed representation $\tilde{\mathbf{H}}_{q}^{e}$. 
The loss enforces high similarity for normal samples while suppressing similarity for anomalous ones, thereby amplifying representation discrepancies associated with abnormal patterns. Formally, the anomaly loss $\mathcal{L}_{\mathrm{AL}}$ for expert training is defined as:
{\small
\begin{equation}
\mathcal{L}\!\left(\mathbf{H}_{q \ i}^{e},\tilde{\mathbf{H}}_{q \ i}^{e},y_i\right)=
\begin{cases}
1-\cos\!\left(\tilde{\mathbf{H}}_{q \ i}^{e},\mathbf{H}_{q \ i}^{e}\right), &if \  y_i=0,\\
\max\!\left(0,\cos\!\left(\tilde{\mathbf{H}}_{q \ i}^{e},\mathbf{H}_{q \ i}^{e}\right)\right), & if \ y_i=1.
\end{cases}
\end{equation}
\begin{equation}
\label{aleq}
\mathcal{L}_{\mathrm{AL}}\!\left(\mathbf{H}_{q }^{e},\tilde{\mathbf{H}}_{q }^{e},\mathbf{y}\right)=
\frac{1}{N}
\sum_{i=1}^{N}
\mathcal{L}\!\left(\mathbf{H}_{q \ i}^{e},\tilde{\mathbf{H}}_{q \ i}^{e},y_i\right),
\end{equation}
}
where $y_i$ denotes the anomaly label of node $i$.


With pretrained experts fixed, the router is trained to learn appropriate expert weights for each sample by introducing a Kullback–Leibler (KL) divergence objective. Specifically, we construct an expert label matrix $\mathbf{Q} \in \{0,1\}^{N \times E}$, where $\mathbf{Q}_{n,e} = 1$ if the pretrained expert $\mathrm{GNN}_e$ correctly predicts node $n$, and $0$ otherwise. The routing logits $\mathbf{G} \in \mathbb{R}^{N \times E}$ are then aligned with $\mathbf{Q}$ through KL divergence:
\begin{equation}
\label{eqkld}
\mathcal{L}_{KL}(\mathbf{Q},\mathbf{G})=\frac{1}{N}\sum_{n=1}^{N}\mathbf{Q}_n\bigl(\log\mathbf{Q}_n-\log\mathbf{G}_n\bigr).
\end{equation}
This formulation explicitly aligns routing predictions with expert performance, leading to stable and effective router optimization.

\subsection{LLM-Based Router Feature Generation}
\label{sec:llm}

Effective expert routing in MoE requires informative and transferable routing signals that capture domain-invariant characteristics.
However, raw node attributes and learned embeddings are often dataset-specific and semantically inconsistent across domains, hindering effective routing learning and limiting robustness in zero-shot scenarios.
To address these limitations, we learn the routing weights $\mathbf{P}$ in Eq.~(\ref{eq:expert_aggregation}) using additional router features.

Specifically, we construct a structured router feature set composed of interpretable graph descriptors, together with an operator set that enables expressive feature generation. We describe router feature set and operation set below, while the full details of both sets are provided in \textbf{Appendix ~\ref{appdenix:router}}.

\noindent\textbf{Router Feature Set.}
We define 23 router features across five categories to capture complementary structural perspectives: (i) PageRank centrality, (ii) betweenness centrality, (iii) closeness centrality, (iv) neighborhood similarity measures, and (v) multi-scale relational summaries.
Each category contains multiple standardized feature primitives computed at the node, ego-graph, and global graph levels, reflecting both local and global structural properties.
The resulting feature set is denoted as $\mathcal{F}$.

\noindent\textbf{Operation Set.}
To enable compositional feature synthesis, we define an operation set $\mathcal{O}$ consisting of three classes of operators:
(i) unary operators that apply nonlinear or monotonic transformations to a single feature (e.g., logarithm, square root, power, sigmoid);  
(ii) binary operators that model relative relationships between two features (e.g., difference, ratio, normalized contrast); and  
(iii) multi-input aggregation operators that summarize multiple features (e.g., mean, variance, sum).  

Building upon sets $\mathcal{F}$ and $\mathcal{O}$, we leverage a large language model (LLM) to automatically synthesize expressive router features through a structured four-stage generation process:
\begin{itemize}[leftmargin=*]
\item The LLM selects one feature category from the five defined in $\mathcal{F}$.
\item  The LLM chooses $k$ features from $\mathcal{F}$ within the selected category: $\{f_1,\dots,f_k\}\subset\mathcal{F}$.
\item The LLM selects an operator $o\in\mathcal{O}$ based on the selected features and subset cardinality: unary for $k=1$ (e.g., logarithm), binary for $k=2$ (e.g., difference), and multi-input aggregation for $k>2$ (e.g., mean).
\item A new combinatorial feature $f'$ is constructed by applying the operator on these $k$ features: $f' = o(f_1,\dots,f_k)$.
\end{itemize}

We adopt an LLM for feature and operator selection instead of predefined random functions, as LLMs provide stronger semantic understanding and reasoning capabilities. By jointly selecting features and conditioning operators on them, the LLM generates semantically meaningful transformations that yield adaptive and diverse router features.

Consequently, this four-stage process is repeated $m$ times to progressively generate diverse router features, which are appended to the feature set via $\mathcal{F} \leftarrow \mathcal{F} \cup \{f'_i\}_{i=1}^{m}$.

\subsection{Router Feature Selection via Efficient Shapley Estimation}
\label{sec:rout}
Despite their enhanced expressiveness, LLM-generated router features may introduce redundant or detrimental dimensions, motivating principled feature contribution assessment and filtering.
To this end, we adopt the Shapley value \cite{rozemberczki2022shapley,fryer2021shapley} from cooperative game theory as a principled metric for evaluating and filtering router features. 
Compared with alternative means (e.g., unrolling the learned model weight assigned to each feature \cite{zhou2022feature}) that are prone to noisy inputs and spurious feature correlations \cite{krell2025influence}, Shapley value provides a robust measurement of feature importance by calculating the average marginal performance gain when each feature is paired with all possible feature subsets.

Specifically, the Shapley value for the feature $f_i$ is computed as:
\begin{equation}
\phi_i
=
\sum_{S \subseteq \mathcal{F} \setminus \{f_i\}}
\frac{|S|!(|\mathcal{F}|-|S|-1)!}{|\mathcal{F}|!}
\bigl[v(S \cup \{f_i\}) - v(S)\bigr],
\end{equation}
where the $\mathcal{F}$ is feature set, and $S$ denotes a subset sampled from $\mathcal{F}\setminus\{f_i\}$.  $v(S)$ denotes the routing utility achieved using feature subset $S$, which will be specified later. The marginal contribution $\Delta_i=v(S \cup \{f_i\}) - v(S)$ captures the performance improvement brought by feature $f_i$ under context $S$. The weighting factor ensures unbiased averaging across all feature orderings. 
A positive $\phi_i$ indicates that feature \( f_i \) provides complementary information beyond \( S \), while a small or negative value suggests a limited or redundant contribution under that feature context.

However, exact computation is prohibitively expensive, as it requires evaluating all possible subsets, resulting in exponential complexity in the number of features.
To efficiently approximate Shapley values, we adopt a complementary subset sampling strategy as shown in Algorithm \ref{alg:shapley_sampling}. 
At each iteration, the feature set is randomly split into two subsets ($S_1$ and $S_2$), enabling simultaneous estimation of marginal contributions for all features via complementary swaps, as shown in lines 5 and 8. 
This approach reduces the computational cost from exponential in $|\mathcal{F}|$ to linear per iteration, as the utility values $v(S_1)$ and $v(S_2)$ are reused to compute marginal contributions for features in the other subsets.

Given $T$ marginal contributions $\{\Delta_{f_i}^{(t)}\}_{t=1}^{T}$ for feature $f_i$, 
we perform significance-based filtering to remove noisy and harmful router features. 
The average contribution is estimated as: $\hat{\phi}_i=\frac{1}{T}\sum_{t=1}^{T}\Delta_{f_i}^{t}$. 
We further assess statistical reliability using the $z$-score: $z_{i}=\frac{\hat{\phi}_{i}}{\mathrm{SE}_{i}}$,
where $\mathrm{SE}_i=\sigma_i/\sqrt{T}$ denotes the standard error and $\sigma_i$ is the standard deviation of  \( \{\Delta_{f_i}^{t}\}_{t=1}^{T} \).  
Finally, the selected feature set is defined as:
\begin{equation}
\mathcal{F}^{+}=\{f_i \in \mathcal{F} \mid \hat{\phi}_i \ge 0 \;\text{or}\; z_i \ge z_{\mathrm{crit}}\},
\end{equation}
where $z_{\mathrm{crit}}$ controls the strictness of the filtering criterion.

\noindent\textbf{Routing Utility $v$.}
The router utility reflects the effectiveness and quality of routing features. A straightforward method to define this utility is to retrain the full MoE model with each newly generated feature set and use the resulting task performance as the routing utility. However, this approach is computationally prohibitive, even without exhaustively evaluating all candidate features. To address this challenge, we leverage the pretrained experts and our proposed KL divergence objective in Eq.~(\ref{eqkld}) to efficiently approximate the routing utility for each feature subset.
Given the router feature subset $\mathcal{S}$ and its router logits $\mathbf{G}^{\mathcal{S}}$, the routing utility is defined as $v(\mathcal{S})=\mathcal{L}_{KL}(\mathbf{Q},\mathbf{G}^{\mathcal{S}})$.
This method quantifies how well the router’s expert assignments align with expert correctness under the given feature subset, thereby efficiently assessing routing feature quality without requiring full MoE retraining.

The detailed time complexity analysis of our efficient Shapley estimation is provided in \textbf{Appendix~\ref{appendix:shapley_time}}.

\begin{algorithm}[t]
\caption{Efficient Marginal Contribution Estimation}
\label{alg:shapley_sampling}
\begin{algorithmic}[1]
    \State\textbf{Input:} Feature set \( \mathcal{F} \), number of iterations \( T \)
    \State\textbf{Output:} Marginal contribution \( \{\Delta_{f_i}^{t}\}_{t=1}^{T} \) for each feature \( f_i \)
\For{\( t = 1 \) to \( T \)}
    \State Randomly partition $\mathcal{F}$ into two subsets $S_1$ and $S_2=\mathcal{F}\setminus S_1$
    \State Compute $v(S_1)$ and $v(S_2)$
    \For{each feature \( f_i \in S_2 \)}
        \State Compute $\Delta_{f_i}^{t} = v(S_1 \cup \{f_i\}) - v(S_1)$
    \EndFor
    \For{each feature \( f_i \in S_1 \)}
        \State Compute $\Delta_{f_i}^{t} = v(S_2 \cup \{f_i\}) - v(S_2)$
    \EndFor
\EndFor
\State Collect the marginal contributions for $f_i \in \mathcal{F}$: \( \{\Delta_{f_i}^{t}\}_{t=1}^{T} \) 
\end{algorithmic}
\end{algorithm}

\subsection{Memory-enhanced Router}
\label{sec:mem}

Despite the Shapley estimation providing a principled criterion for identifying informative routing features, its effectiveness fundamentally depends on the effectiveness of the router, while the initially constructed features are limited in expressiveness as they are generated through single-step operations. To jointly improve feature quality and selection reliability, we therefore design an iterative feature generation framework, as illustrated in the top panel of Fig.~\ref{fig:framework}. After each round, the router is retrained on the updated feature set to produce more accurate routing weights, which in turn enable more reliable feature evaluation and selection. 
This closed-loop process progressively enhances both feature quality and routing capability, subject to the requirement that routing-relevant knowledge is accumulated across successive rounds.



To this end, we introduce two memory banks within the MoE router $\mathcal{R}$ to store \textbf{domain invariant routing principal} and enhance the final routing weight $\mathbf{P}$:
\begin{equation}
\label{eq:rot}
\mathbf{P} = \mathcal{R}\!\left(\mathbf{H}_r, \mathcal{G}\right).
\end{equation}
Specifically, we first project node features $\tilde{\mathbf{X}}\in\mathbb{R}^{N\times d}$ and router features $\mathbf{H}_r\in\mathbb{R}^{N\times d_r}$ into query embeddings:
\begin{equation}
\label{eq:linear_projection}
\mathbf{H}_n^{q}=\mathrm{GNN}_r(\tilde{\mathbf{X}},\mathbf{A}), \qquad
\mathbf{H}_r^{q} =\mathbf{H}_r\mathbf{W}_r+\mathbf{b}_r,
\end{equation}
where $\mathbf{W}_r\in\mathbb{R}^{d_r\times d_m}$ and $\mathbf{b}_r\in\mathbb{R}^{d_m}$ are learnable parameters. $d_r$ and $d_m$ denote the router feature and memory dimensions.
These query embeddings are then used to retrieve relevant memory embeddings:
\begin{equation}
\label{eq:query_memory}
\mathbf{S}_n=\mathrm{softmax}(\mathbf{H}_n^{q}\boldsymbol{\Phi}_n^{\top}),\quad 
\mathbf{S}_r=\mathrm{softmax}(\mathbf{H}_r^{q}\boldsymbol{\Phi}_r^{\top})
\end{equation}
\begin{equation}
\mathbf{H}_n^{m}=\mathbf{S}_n\boldsymbol{\Phi}_n,\quad \mathbf{H}_r^{m}=\mathbf{S}_r\boldsymbol{\Phi}_r,
\end{equation}
where $\boldsymbol{\Phi}_n,\boldsymbol{\Phi}_r\in\mathbb{R}^{M\times d_m}$ denote the memory banks for node and router embeddings, respectively, and $M$ is the memory size.

The retrieved $\mathbf{H}_n^{m}$ and $\mathbf{H}_r^{m}$ are used to compute routing weights for expert selection. 
For node $i$, the routing logit for expert $e$ is obtained by applying expert-specific feature scaling followed by similarity estimation:
\begin{equation}
g_i^{e}=\mathrm{sim}\!\left(\mathbf{H}_{r \ i}^{m},\,\mathbf{H}_{n \ i}^{m}\odot\mathbf{s}^{e}\right),
\qquad e=1,\dots,E,
\end{equation}
where $\mathbf{s}^{e}\in\mathbb{R}^{d_m}$ is the learnable scaling vector and $\mathrm{sim}(\cdot,\cdot)$ denotes dot-product similarity.
By stacking $\{g_i^{e}\}_{e=1}^{E}$, we obtain the routing logit vector $\mathbf{g}_i\in\mathbb{R}^{E}$, which reflects the affinity between node $i$ and each expert. 
To encourage exploration of experts, we adopt a noise term \cite{shazeer2017outrageously,wang2023graph} to compute routing weights:
\begin{equation}
\label{eq:noisy_routing}
\mathbf{G}_i=\mathbf{g}_i+\epsilon\cdot\mathrm{Softplus}(\mathbf{H}_{r \ i}\mathbf{W}),\qquad
\mathbf{P}_i=\mathrm{Softmax}(\mathbf{G}_i),
\end{equation}
where $\epsilon$ is Gaussian noise and $\mathbf{W}$ is a learnable parameter matrix. \(\mathbf{G}\in\mathbb{R}^{N\times E}\) is the router logits and the resulting routing weight $\mathbf{P}\in\mathbb{R}^{N\times E}$ encodes per-node expert assignment probabilities.

\subsection{Invariant Routing Learning}

The generated router features substantially enrich routing signals but may also introduce spurious correlations that vary across graph domains.
As a result, the router may overfit to domain-specific feature patterns rather than capturing fundamental relationships between graph structures and expert behaviors.
To address this issue, we adopt the invariant learning principle that encourages the router to focus on stable, domain-independent routing patterns while suppressing spurious correlations.

Specifically, we generate $K$ augmented router features via masking on \( \mathbf{H}_r \):
\begin{equation}
\label{eq:random_feature_mask}
\mathbf{H}^{k}_r=\mathbf{H}_r \odot \mathbf{m}^{k},\quad k = 1, \dots, K,
\end{equation}
where \( \mathbf{m}^{k} \) is a randomly sampled binary feature mask, and \( \mathbf{H}^{k}_r \) represents the masked router features under the \( k \)-th environment.
For each masked router feature \( \mathbf{H}^{k}_r \), we feed it into the router $\mathcal{R}$ to obtain the corresponding routing weights $\mathbf{P}^{k}$ via Eq. (\ref{eq:rot}).
Conditioned on the different routing weights \( \mathbf{P}^{k} \), we get the corresponding the expert-specific query embeddings and their reconstruction embeddings similar to Eq. (\ref{eq:expert_aggregation}):
\begin{equation}
\label{eq:expert_aggregation2}
\mathbf{H}_q^{k}=\sum_{e=1}^{E} \mathbf{P}^k_{:,e}\mathbf{H}_q^{e}, \quad
\tilde{\mathbf{H}}_q^{k}=\sum_{e=1}^{E} \mathbf{P}^k_{:,e}\tilde{\mathbf{H}}_q^{e},
\end{equation}
For each environment $k$, we first compute the router training loss 
$\mathcal{L}^k_{\mathrm{A}}=\mathcal{L}_{\mathrm{AL}}\!\left(\mathbf{H}_q^{k},\tilde{\mathbf{H}}_q^{k},\mathbf{y}\right)$ according to Eq.~(\ref{aleq}). 
Then, we construct the invariant objective $\mathcal{L}_{\mathrm{IN}}$ by minimizing both the mean risk and its variability:
\begin{equation}
\mathcal{L}_{\mathrm{IN}}
=
\frac{1}{K}\sum_{k=1}^{K}\mathcal{L}_{\mathrm{A}}^{k}
+
\lambda\,\mathrm{Var}\!\left(\{\mathcal{L}_{\mathrm{A}}^{k}\}_{k=1}^{K}\right),
\end{equation}
where $K$ denotes the number of environments, $\lambda$ is a balancing hyperparameter, and
$\mathrm{Var}(\cdot)$ computes the variance across environments.

The first term minimizes the average training error, while the variance regularizer penalizes environment-dependent performance fluctuations, thereby encouraging the router to learn stable and invariant routing patterns.

In addition, to regularize expert utilization and avoid routing collapse, we incorporate standard MoE balancing losses that encourage uniform expert engagement. 
Specifically, we penalize the coefficient of variation (CV) \cite{abdi2010coefficient} of both the aggregated routing weights $P_{i,e}$ and routing logits $G_{i,e}$ across experts:
\begin{equation}
\label{eq:aux_losses}
\mathcal{L}_{\mathrm{MoE}}
=
\mathrm{CV}\!\left(\sum_{i=1}^{N} P_{i,e}\right)^2
+
\mathrm{CV}\!\left(\sum_{i=1}^{N} G_{i,e}\right)^2 ,
\end{equation}
This regularization encourages balanced gating magnitudes and computational load across experts during training~\cite{wang2023graph,shazeer2017outrageously}. The overall loss for router training is:
\begin{equation}
\label{overloss}
\mathcal{L}_{\mathrm{final}}=\mathcal{L}_{\mathrm{IN}}+\mathcal{L}_{\mathrm{MoE}}.
\end{equation}
Finally, the learned domain invariant routing principles are stored in the router memory, enabling consistent and transferable expert allocation across unseen graph domains.

\begin{algorithm}[t]
\caption{Overall Training Pipeline}
\label{alg:pp}
\begin{algorithmic}[1]
\State \textbf{Input:} Graph $\mathcal{G}$, expert GNNs, router $\mathcal{R}$, LLM.
\State \textbf{Output:} Trained router $\mathcal{R}$ and experts.

\State Pre-train expert GNNs using Eq.~(\ref{aleq}).
\State Warm up $\mathcal{R}$ using Eq.~(\ref{eqkld}).
\State Initialize router features $\mathcal{F}$ and operators $\mathcal{O}$.

\For{$r=1$ to $R$}
    \State Generate $m$ features: $\mathcal{F}\!\leftarrow\!\mathcal{F}\cup\{f'_i\}_{i=1}^{m}$.
    \State Select features via Shapley estimation and $\mathcal{F}\!\leftarrow\!\mathcal{F}^{+}$.
    \State Update $\mathcal{R}$ by Eq.~(\ref{overloss}) with experts frozen.
\EndFor

\State Retrain $\mathcal{R}$ with final $\mathcal{F}$

\end{algorithmic}
\end{algorithm}

\begin{table*}[t]
\setlength{\abovecaptionskip}{1pt}
\renewcommand{\arraystretch}{0.9}
\centering
\small
\setlength{\tabcolsep}{6pt}
\caption{Zero-shot GAD performances on nine real-world datasets. 
The best performance per column is boldfaced, and the second-best is underlined. 
“Rank” indicates each method's average rank over all datasets.}
\label{tab:zero_shot_gad_auc_prc}
\begin{tabular}{c|c|c|cccccc|c}
\toprule
\multirow{2}{*}{\textbf{Metric}} 
& \multirow{2}{*}{\textbf{Category}} 
& \multirow{2}{*}{\textbf{Method}} 
& \multicolumn{6}{c|}{\textbf{Dataset}} 
& \multirow{2}{*}{\textbf{Rank}} \\
\cmidrule(lr){4-9}
& & 
& Cora 
& CiteSeer 
& ACM 
& BlogCatalog 
& Facebook 
& Weibo 
&  \\
\midrule

\multirow{9}{*}{\rotatebox{90}{\text{AUROC}}}
& \multirow{5}{*}{\begin{tabular}{c}
Supervised\\
method
\end{tabular}}

& SAGE        & 43.35 ± 2.79 & 38.98 ± 3.71 & \underline{83.28 ± 0.31} & 57.48 ± 3.67 & \textbf{73.38 ± 5.54} & 61.06 ± 2.87 & 5.83 \\
& & GIN        & 56.82 ± 0.26 & 57.65 ± 0.31 & 42.68 ± 0.44 & 56.62 ± 0.83 & 39.71 ± 1.91 & 73.31 ± 3.17  & 6.83 \\
& & BWGNN      & 62.60 ± 0.19 & 67.72 ± 1.23 & 80.54 ± 1.11 & \underline{78.00 ± 0.25} & 63.16 ± 0.40 & 63.24 ± 2.81  & 3.83 \\
& & AMNet      & 58.26 ± 1.00 & 60.88 ± 2.55 & 74.92 ± 1.39 & 69.47 ± 0.83 & 63.65 ± 2.52 & 59.76 ± 6.29  & 5.67 \\
& & GHRN       & 60.17 ± 0.94 & 61.53 ± 0.32 & 75.73 ± 0.18 & \textbf{78.52 ± 0.22} & \underline{69.87 ± 1.13} & 71.52 ± 0.23  & 3.50 \\

\cline{2-3}
& \multirow{4}{*}{\begin{tabular}{c}
Generalist\\
method
\end{tabular}}
& ARC        & \underline{83.98 ± 0.27} & \underline{90.82 ± 0.36} & 77.29 ± 0.11 & 74.98 ± 0.26 & 63.74 ± 0.30 & {89.76 ± 0.20}  & 3.17 \\
& & UNPrompt   & 55.08 ± 0.84 & 55.84 ± 1.03 & 69.31 ± 0.68 & 68.71 ± 0.44 & 56.63 ± 5.43 & 53.88 ± 1.65  & 7.17 \\
& & AnomalyGFM & 53.39 ± 1.08 & 53.26 ± 0.22 & 55.18 ± 3.43 & 54.19 ± 0.26 & 32.21 ± 3.23 & \underline{92.09 ± 1.74}  & 7.33 \\
& & EvoFG     & \textbf{84.27 ± 0.34} & \textbf{91.57 ± 0.03} & \textbf{91.62 ± 1.42} & 75.11 ± 0.47 & 66.26 ± 3.24& \textbf{93.05 ± 0.88} 
& \textbf{1.67} \\

\midrule
\midrule
\multirow{9}{*}{\rotatebox{90}{\text{AUPRC}}}
& \multirow{5}{*}{\begin{tabular}{c}
Supervised\\
method
\end{tabular}}
& SAGE        & 5.56 ± 0.38  & 4.72 ± 0.86  & 23.62 ± 0.87  & 15.49 ± 3.74 & \textbf{8.70 ± 3.10}  & 36.20 ± 4.30 & 6.17 \\
& & GIN        & 6.22 ± 0.06  & 5.68 ± 0.09  & 6.90 ± 0.27  & 23.35 ± 0.36 & 1.85 ± 0.08  & 20.44 ± 2.64  & 7.33 \\
& & BWGNN      & 10.31 ± 0.20 & 16.06 ± 0.83 & 30.05 ± 0.86 & \textbf{37.39 ± 0.25} & 5.65 ± 1.71  & 13.61 ± 0.91  & 4.00 \\
& & AMNet      & 8.50 ± 0.69  & 8.08 ± 1.37  & 29.23 ± 1.59 & 31.77 ± 0.96 & 5.13 ± 0.38  & 13.28 ± 2.10  & 5.67 \\
& & GHRN       & 12.04 ± 0.81 & 12.02 ± 1.19 & 24.43 ± 0.60 & \underline{36.89 ± 0.09} & \underline{6.22 ± 0.71}  & 21.78 ± 0.69  & 3.67 \\
\cline{2-3}
& \multirow{4}{*}{\begin{tabular}{c}
Generalist\\
method
\end{tabular}}
& ARC        & \underline{45.42 ± 1.30} & \underline{47.79 ± 1.75} & \underline{39.41 ± 0.04} & 36.64 ± 0.23 & 5.94 ± 0.21  & \underline{69.15 ± 0.39}  & 2.50 \\
& & UNPrompt   & 7.65 ± 1.08  & 5.74 ± 0.41  & 14.53 ± 2.91 & 24.82 ± 1.97 & 2.93 ± 0.45  & 26.89 ± 1.93  & 6.17 \\
& & AnomalyGFM & 5.97 ± 0.16  & 4.88 ± 0.32  & 4.60 ± 0.30  & 7.13 ± 0.39  & 1.63 ± 0.08  & 59.62 ± 2.03  & 7.67 \\
& & EvoFG     & \textbf{47.71 ± 0.99} & \textbf{52.09 ± 1.78} & \textbf{45.78 ± 1.43} & 36.51 ± 0.34 & 6.09 ± 0.87  & \textbf{69.80 ± 1.59} & \textbf{1.83} \\
\bottomrule
\end{tabular}
\vspace{-0.3cm}
\end{table*}

\subsection{Overall Training Pipeline}
\label{sec:pipeline}
The overall training procedure of the proposed framework is presented in Algorithm~\ref{alg:pp}. We first pre-train all expert GNNs independently on the training datasets and then warm up the router. Subsequently, router feature generation is performed for $R$ iterative rounds. At each round, an LLM generates $m$ new router features conditioned on the current feature set $\mathcal{F}$.
These features are then evaluated through the efficient Shapley estimation, producing a refined feature set $\mathcal{F}^{+}$.
To accommodate the updated feature dimensionality, the router's linear projection layer in Eq.(\ref{eq:linear_projection}) is reinitialized, and the router is retrained with expert GNNs frozen. By iteratively constructing semantically richer features on top of previously generated ones and training the memory-enhanced router via invariant risk minimization, we can largely enhance the routing efficacy in the MoE for zero-shot generalist GAD.

\section{Experiments}
We design comprehensive experiments to validate the effectiveness of EvoFG and aim to answer the following questions. 
\noindent\textbf{Q1}: Can EvoFG achieve better zero-shot performance then baselines?
\textbf{Q2}: How do the different components affect EvoFG?
\textbf{Q3}: How do the different hyper-parameters affect EvoFG?
\textbf{Q4}: How does the router behave in terms of expert selection?
\textbf{Q5}: How does each expert perform in EvoFG?
Due to space limitations, we delay \textbf{Q5}, and some additional results for \textbf{Q4} to \textbf{Appendix \ref{appendix:e}}.

\subsection{Experimental Setup}
\label{secsetting}
\noindent\textbf{Datasets.} 
We evaluate EvoFG on diverse graph datasets across citation, social, and e-commerce domains.
To train and evaluate all models under strict zero-shot setting, we adopt a cross-dataset evaluation protocol \cite{qiao2025anomalygfm}.
Specifically, all models are trained on PubMed~\cite{sen2008collective} and Flickr~\cite{ding2019deep, tang2009relational}, validated on Coauthor-CS~\cite{dong2025g} and Amazon-Photo~\cite{shchur2018pitfalls}, and finally tested on BlogCatalog~\cite{ding2019deep, tang2009relational}, Facebook~\cite{xu2022contrastive}, Weibo~\cite{kumar2019predicting}, Cora~\cite{tang2008arnetminer}, CiteSeer~\cite{tang2008arnetminer}, and ACM~\cite{sen2008collective}. The full dataset statistics are outlined in  \textbf{Appendix~\ref{appendix:datasets}}.


\noindent\textbf{Baselines.} We compare EvoFG with conventional GNNs (SAGE~\cite{hamilton2017inductive}, GIN~\cite{xu2018powerful}), state of the art GAD methods (BWGNN~\cite{tang2022rethinking}, AMNet~\cite{chai2022can}, and GHRN~\cite{gao2023addressing}), zero shot approaches (UNPrompt~\cite{niu2024zero} and AnomalyGFM~\cite{qiao2025anomalygfm}), and few shot model ARC~\cite{liu2024arc}, where in context examples are drawn from training data for zero shot evaluation. We provide further descriptions of all baselines in \textbf{Appendix~ \ref{appendix:baselines}}.

\noindent\textbf{Implementation Notes.} EvoFG uses Qwen2-7B~\cite{team2024qwen2} as the LLM backbone. In EvoFG, we adopt four GNN experts, namely GCN \cite{kipf2016semi}, GAT \cite{velivckovic2018graph}, ChebNet \cite{defferrard2016convolutional}, and GPR \cite{chien2020adaptive}. The full details of GNN experts, hyperparameters, and hardware are provided in \textbf{Appendix~\ref{appendix:Implementation}}.

\noindent\textbf{Metrics.}
Following prior works~\cite{tang2023gadbench,huang2022hop,liu2024arc,qiao2025anomalygfm}, we employ AUROC and AUPRC as evaluation metrics and report the mean and standard deviation over three independent runs. 
\vspace{-0.3cm}



\begin{table*}[]
\setlength{\abovecaptionskip}{1pt}
\caption{Ablation study results for EvoFG and its variants. }
\label{tab:ablation}
\resizebox{0.8\textwidth}{!}{
\begin{tabular}{c|l|lllllll}
\toprule
\multirow{2}{*}{\textbf{Metric}} 
& \multicolumn{1}{c|}{\multirow{2}{*}{\textbf{Method}}} 
& \multicolumn{6}{c}{\textbf{Dataset}} & \multicolumn{1}{|c}{\multirow{2}{*}{\textbf{Rank}}} \\
\cmidrule(lr){3-8}
 & \multicolumn{1}{c}{} 
 & \multicolumn{1}{|c}{Cora} 
 & \multicolumn{1}{c}{CiteSeer} 
 & \multicolumn{1}{c}{ACM} 
 & \multicolumn{1}{c}{BlogCatalog} 
 & \multicolumn{1}{c}{Facebook} 
 & \multicolumn{1}{c}{Weibo} & \multicolumn{1}{|c}{}\\
\midrule

\multirow{5}{*}{\rotatebox{90}{AUROC}}
& w RandFeat
& \textbf{85.19 ± 0.27} & \underline{91.69 ± 1.78} & 88.86 ± 3.01 & 75.04 ± 0.67 & \underline{63.17 ± 0.91} & 92.36 ± 0.98 & \multicolumn{1}{|c}{2.50}  \\
& w/o Feature Selection
& 83.07 ± 1.95 & \textbf{91.77 ± 3.01} & 87.36 ± 0.85 & 74.95 ± 0.54 & 63.13 ± 1.48 & 91.97 ± 1.19 &  \multicolumn{1}{|c}{3.33} \\
& w/o Router Memory
& 74.75 ± 16.15 & 66.98 ± 9.97 & \underline{91.05 ± 4.57} & 74.81 ± 4.24 & 55.91 ± 9.15 & 90.49 ± 2.37  &  \multicolumn{1}{|c}{4.50} \\
& w/o Invariant Learning
& 82.43 ± 2.49 & 90.93 ± 0.69 & 90.76 ± 2.89 & \textbf{75.70 ± 0.62} & 61.25 ± 4.96 & \underline{92.41 ± 0.39}  &  \multicolumn{1}{|c}{3.00} \\
& \textbf{EvoFG (Full)}
& \underline{84.27 ± 0.34} & 91.57 ± 0.03 & \textbf{91.62 ± 1.42} & \underline{75.11 ± 0.47} & \textbf{66.26 ± 3.24} & \textbf{93.05 ± 0.88}   &  \multicolumn{1}{|c}{\textbf{1.67}} \\
\midrule
\midrule
\multirow{5}{*}{\rotatebox{90}{AUPRC}}
& w RandFeat
& \textbf{48.30 ± 0.57} & \textbf{53.84 ± 3.46} & 42.88 ± 0.48 & 36.24 ± 0.23 & \underline{5.40 ± 0.39} & \underline{67.02 ± 7.38}  &  \multicolumn{1}{|c}{2.17} \\
& w/o Feature Selection
& 44.50 ± 0.82 & 51.28 ± 5.71 & 42.73 ± 1.31 & 36.05 ± 0.25 & 4.65 ± 0.96 & 66.88 ± 1.62  &  \multicolumn{1}{|c}{3.67}  \\
& w/o Router Memory
& 22.57 ± 10.96 & 15.41 ± 9.78 & 43.85 ± 8.10 & 35.05 ± 3.63 & 4.43 ± 2.33 & 55.76 ± 12.61  &  \multicolumn{1}{|c}{4.50} \\
& w/o Invariant Learning
& 45.23 ± 3.99 & 48.80 ± 4.54 & \underline{44.86 ± 1.72} & \textbf{36.67 ± 0.33} & 4.42 ± 1.01 & 66.76 ± 3.74   &  \multicolumn{1}{|c}{3.17}  \\
& \textbf{EvoFG (Full)}
& \underline{47.71 ± 0.99} & \underline{52.09 ± 1.78} & \textbf{45.78 ± 1.43} & \underline{36.51 ± 0.34} & \textbf{6.09 ± 0.87} & \textbf{69.80 ± 1.59}   &  \multicolumn{1}{|c}{\textbf{1.50}}\\
\bottomrule
\end{tabular}
}
\vspace{-0.3cm}
\end{table*}



\subsection{Zero-shot Performance (Q1)}
From Table~\ref{tab:zero_shot_gad_auc_prc}, we first observe that conventional GNN models such as SAGE and GIN generally exhibit limited performance in the zero-shot GAD setting, as they are not explicitly designed to capture anomaly-related patterns and primarily focus on supervised representation learning. In contrast, GAD-specific models including BWGNN, AMNet, and GHRN achieve strong performance on several datasets—for example, BWGNN on ACM and BlogCatalog, and GHRN on BlogCatalog and Weibo—demonstrating the benefit of incorporating anomaly-aware inductive biases into model design. Among generalist GAD baselines, ARC consistently performs the best, which can be attributed to its use of in-context nodes from the training graphs during inference, providing informative references for anomaly detection in unseen domains. However, other generalist approaches such as UNPrompt and AnomalyGFM show comparatively weaker results, suggesting that the transferable knowledge they learn may not effectively align with the anomaly characteristics of target datasets. Overall, EvoFG achieves the most consistent and superior performance across datasets, as the LLM-generated features enrich the routing space and enable the router to select appropriate experts that better match the structural and semantic properties of each target graph.

\vspace{-0.2cm}
\subsection{Ablation Study (Q2)}
We evaluate several variants of EvoFG to analyze the contribution of each component. 
\textbf{w RandFeat} replaces the LLM-based decision mechanism in the feature evolution process with random selection, while keeping the same feature generation patterns and operators unchanged, 
\textbf{w/o Feature Selection} removes the Shapley-based feature selection strategy, 
\textbf{w/o Router Memory} replaces the memory-enhanced router with a simple fully connected routing layer, 
and \textbf{w/o Invariant Learning} disables the invariant learning objective.

Overall, the full \textbf{EvoFG (Full)} model achieves the most stable and competitive zero-shot GAD performance across all datasets. 
Although some variants outperform the full model on individual datasets, their performance is inconsistent and lacks robustness.

In particular, \textbf{w RandFeat} performs comparably to \textbf{EvoFG (Full)} on Cora and CiteSeer, while showing degraded performance on the remaining datasets.
This can be explained by two factors. First, randomly generated features may occasionally capture dataset-specific signals, but their quality varies significantly across datasets, leading to unstable and biased behavior under zero-shot settings. In contrast, LLM-guided feature evolution leverages prior knowledge and structural awareness, producing more semantically meaningful and transferable router features with more stable performance.
Second, since PubMed (one of the training datasets) shares similar citation-network characteristics with Cora and CiteSeer, random features can generalize moderately well to these datasets, but fail on others with substantially different distributions. LLM-generated features are less tied to dataset-specific distributions and thus exhibit stronger robustness across heterogeneous zero-shot GAD benchmarks.
\textbf{w/o Feature Selection} results in a mild performance drop, indicating that feature selection is beneficial but not strictly essential. 
\textbf{w/o Router Memory} suffers from substantial performance degradation and large variance, suggesting that a simple fully connected router lacks sufficient capacity to accumulate and reuse routing knowledge across iterations. 
\textbf{w/o Invariant Learning} slightly degrades performance and increases variance on certain datasets, highlighting its role in encouraging invariant and transferable routing behavior. 
Overall, each component contributes to the robustness and generalization ability of EvoFG.

\subsection{Hyper-parameter Analysis (Q3)}

\begin{figure}[t]
\setlength{\abovecaptionskip}{1pt}
    \centering
    \includegraphics[width=0.4\textwidth]{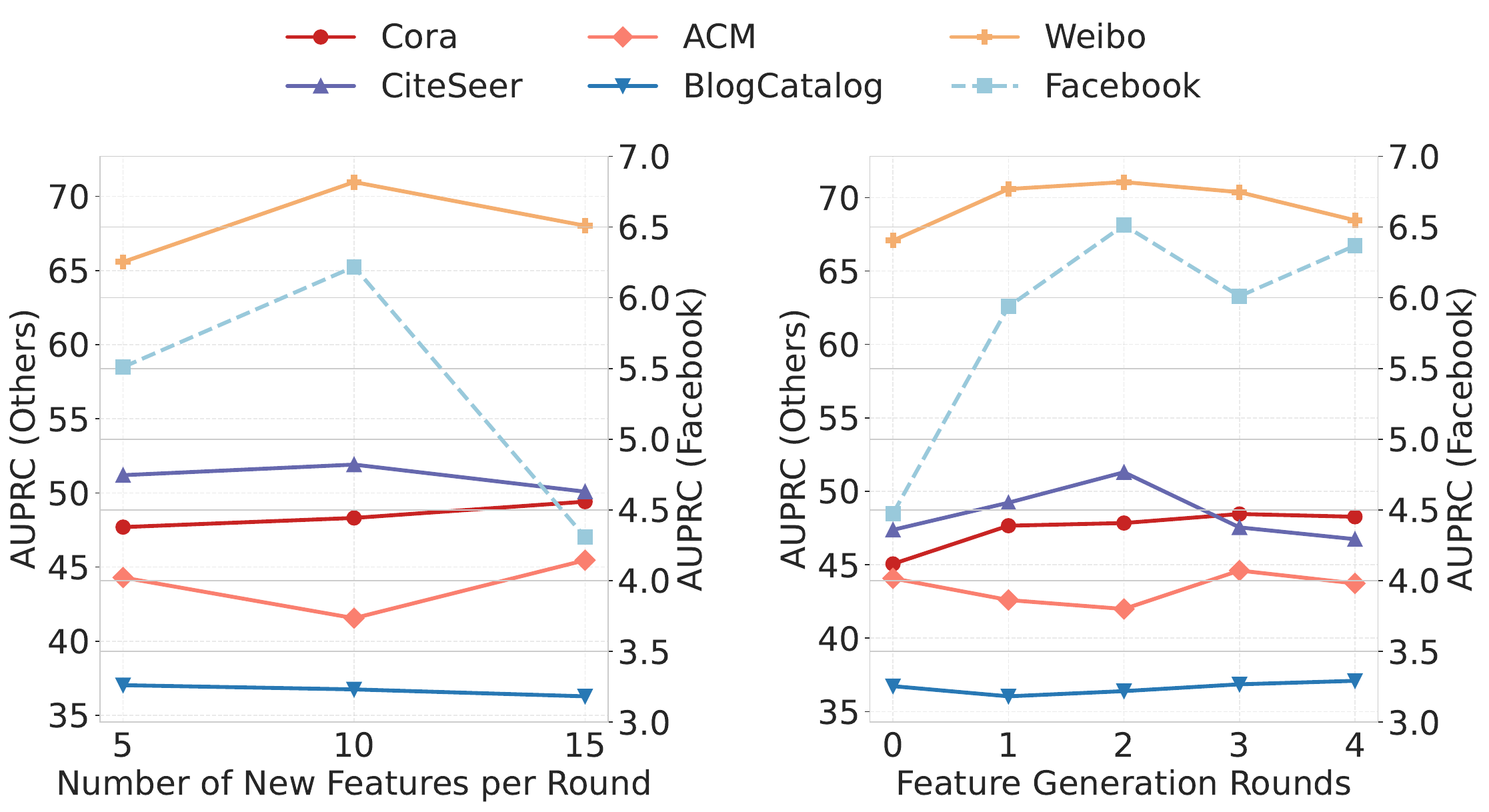}
    \caption{
    Hyper-parameter analysis of feature generation.
    }
    \setlength{\abovecaptionskip}{1pt}
    \label{fig:rounds_vs_new_features_auprc}
    \vspace{-0.7cm}
\end{figure}

We further analyze the sensitivity of our method to two key hyper-parameters in the feature generation process, namely the number of newly generated features per round $m$ and the total number of feature generation rounds $R$, as illustrated in Fig.~\ref{fig:rounds_vs_new_features_auprc}.
As shown in the left panel, increasing the number of new features per round generally improves performance at the early stage, indicating that enriching the feature space helps the router better align samples with appropriate experts. However, overly aggressive feature expansion may introduce redundant or noisy features, leading to performance saturation or slight degradation on some datasets (e.g., CiteSeer and Facebook). In practice, a moderate number of newly generated features per round provides a good trade-off between performance gains and feature quality.

The right panel studies the impact of feature generation rounds. We observe that AUPRC consistently improves in the first few rounds, demonstrating the effectiveness of the iterative LLM-based feature generation mechanism. Nevertheless, after a certain number of rounds, the performance gain becomes marginal and may even fluctuate, suggesting diminishing returns from repeatedly expanding the feature space. 
Overall, the results indicate that EvoFG is robust to a reasonable range of hyper-parameter choices and achieves stable performance with a small number of feature generation rounds and a moderate feature budget.

\begin{figure}[t]
\setlength{\abovecaptionskip}{1pt}
    \centering
    \includegraphics[width=0.7\columnwidth]{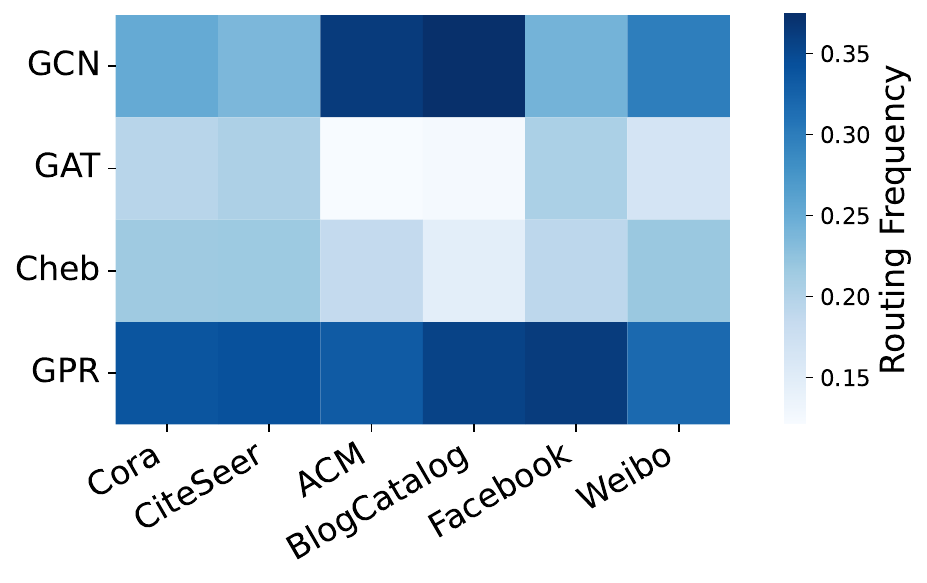}
    \caption{Soft routing heatmap across datasets.}
    \setlength{\abovecaptionskip}{1pt}
    \label{fig:evofg_full_heatmap}
    \vspace{-0.6cm}
\end{figure}

\subsection{Case Study on Routing Behavior (Q4)}
\label{case1}
We visualize the routing behavior of EvoFG using a heatmap of routing frequency, which reflects how often the router assigns each expert across different datasets. This visualization demonstrates that the proposed router is capable of allocating experts adaptively according to dataset-specific distributions. Specifically, the GPR expert is selected with a relatively high frequency across all datasets, indicating its general effectiveness in capturing broadly useful structural patterns. On the ACM and BlogCatalog datasets, GCN is favored more frequently, while GAT is selected with a much lower frequency, suggesting that different graph characteristics lead the router to emphasize different inductive biases. The routing frequency distributions on Cora and CiteSeer are highly similar, which is consistent with the fact that they originate from closely related citation domains. 
Overall, the routing frequencies vary substantially across datasets, and no single expert dominates with an extremely high or low selection probability, indicating that the router avoids expert collapse and generalizes well across different data distributions. 

Additional heatmaps for ablation variants and the formal definition of routing frequency are provided in the \textbf{Appendix~\ref{appendix:selection}}.

\section{Conclusion}

To our knowledge, this is the first work to enhance graph anomaly detection generalization by explicitly targeting the MoE router. Rather than strengthening individual experts, we reveal the router as a key bottleneck for cross dataset transfer and propose a unified framework that improves routing generalization via LLM based feature generation and selection. A memory augmented router and invariant learning paradigm further enable stable and transferable expert assignment across heterogeneous graphs. Extensive experiments demonstrate robust routing behavior and superior generalization performance under distribution shifts.








\begin{acks}
To Robert, for the bagels and explaining CMYK and color spaces.
\end{acks}

\bibliographystyle{ACM-Reference-Format}
\bibliography{sample-base}

\newpage
\appendix
\begin{center}
\Large \bf {Appendix}
\end{center}
\etocdepthtag.toc{mtappendix}
\etocsettagdepth{mtchapter}{none}
\etocsettagdepth{mtappendix}{subsection}
\tableofcontents

\section{Dataset Description}
\label{appendix:datasets}
\noindent\textbf{Training Datasets.}
The training stage is conducted on two large-scale graphs that provide rich structural and semantic information.

\textbf{PubMed}~\cite{sen2008collective} is a citation network in which nodes correspond to academic publications and edges denote citation relationships between papers. Each node is associated with a bag-of-words feature vector extracted from the textual content of the publication.

\textbf{Flickr}~\cite{ding2019deep, tang2009relational} is a social network dataset that models user interactions in a photo-sharing platform. Users are treated as nodes, while edges represent mutual following relationships. Node attributes are derived from user-generated content, such as textual tags associated with shared photos.

These two datasets serve as the training graphs for learning expert representations and routing behaviors, as they differ substantially in both domain and graph structure.

\vspace{0.5em}
\noindent\textbf{Validation Datasets.}
To select hyper-parameters and validate model performance during training, we employ two datasets from distinct domains that are disjoint from the training data. These datasets are used exclusively for validation and model selection, ensuring that no information from the test graphs is used during training.

\textbf{Coauthor-CS}~\cite{dong2025g} is a collaboration network extracted from computer science publication records. Nodes represent authors, and edges indicate co-authorship relationships. Node features are constructed as bag-of-words vectors derived from paper titles and abstracts, reflecting the research topics of individual authors.

\textbf{Amazon Photo}~\cite{shchur2018pitfalls} is a co-purchase network derived from the Amazon platform, where nodes correspond to products and edges indicate that two products are frequently purchased together. Node attributes are extracted from product descriptions and metadata, forming high-dimensional feature representations.

\vspace{0.5em}
\noindent\textbf{Test Datasets.}
The proposed method is finally evaluated on a diverse set of unseen graphs spanning multiple domains, including citation networks and social networks.

\textbf{Cora}, \textbf{CiteSeer}, and \textbf{ACM}~\cite{tang2008arnetminer} are citation network datasets, where nodes correspond to academic publications and edges represent citation relationships. Each publication is represented by a bag-of-words feature vector, with the dimensionality determined by the size of the corresponding vocabulary.

\textbf{BlogCatalog}~\cite{ding2019deep, tang2009relational} is a social network dataset that models user interactions in an online blogging platform. Users are represented as nodes, and edges capture mutual following relationships. Node attributes are derived from textual content generated by users.

\textbf{Facebook}~\cite{xu2022contrastive} is a social network dataset that captures friendship relations among users, where nodes represent individuals and edges encode social connections formed through friend interactions.

\textbf{Weibo}~\cite{kumar2019predicting} is a social media dataset constructed from Tencent Weibo, modeling user activities and hashtag usage. Users who exhibit frequent suspicious posting behaviors within a short temporal window are labeled as anomalies. Node attributes consist of location information and bag-of-words features extracted from microblog posts.

These datasets are used solely for testing, enabling a comprehensive evaluation of the model’s zero-shot generalization performance across different graph domains and anomaly characteristics.

\begin{table}[t]
\centering
\footnotesize
\setlength{\tabcolsep}{2pt}
\caption{Statistics of datasets used in this work.}
\label{tab:dataset_statistics}
\begin{tabular}{l l rrrrr}
\toprule
\textbf{Split} 
& \textbf{Dataset} 
& \textbf{\#Nodes} 
& \textbf{\#Edges} 
& \textbf{\#Feat.} 
& \textbf{Avg. Deg.} 
& \textbf{\#Anomalies} \\
\midrule
\multirow{2}{*}{\textbf{Training}} 
& PubMed      & 19,717 & 44,338  & 500    & 4.50  & 600 (3.04\%) \\
& Flickr      & 7,575  & 239,738 & 12,047 & 63.30 & 450 (5.94\%) \\
\midrule
\multirow{2}{*}{\textbf{Validation}} 
& Coauthor-CS & 18,333 & 163,788 & 6,805  & 17.86 & 600 (3.27\%) \\
& Amazon-Photo & 7,650 & 238,162 & 745    & 62.30 & 450 (5.55\%) \\
\midrule
\multirow{6}{*}{\textbf{Test}} 
& Cora        & 2,708  & 5,429   & 1,433  & 3.90  & 150 (5.53\%) \\
& CiteSeer    & 3,327  & 4,732   & 3,703  & 2.77  & 150 (4.50\%) \\
& ACM         & 16,484 & 71,980  & 8,337  & 8.73  & 597 (3.62\%) \\
& BlogCatalog & 5,196  & 171,743 & 8,189  & 66.11 & 300 (5.77\%) \\
& Facebook    & 1,081  & 55,104  & 576    & 50.97 & 25 (2.31\%) \\
& Weibo       & 8,405  & 407,963 & 400    & 48.53 & 868 (10.30\%) \\

\bottomrule
\end{tabular}
\end{table}


\section{Baselines Description}
\label{appendix:baselines}
A more detailed introduction of the GAD models we compare with is given as follows.

\textbf{SAGE}~\cite{hamilton2017inductive} applies neighborhood aggregation via message passing, where each node updates its representation by aggregating normalized features from its neighbors. It serves as a widely used backbone for graph representation learning.

\textbf{GIN}~\cite{xu2018powerful} employs the Graph Isomorphism Network, which uses sum-based neighborhood aggregation followed by multi-layer perceptrons to achieve strong discriminative power in graph representation learning. In GAD settings, GIN serves as a representative expressive GNN backbone for modeling complex local structures.

\textbf{BWGNN}~\cite{tang2022rethinking} is a frequency-aware GAD method that models the spectral energy distribution of graphs and introduces a localized band-pass filter to capture informative high-frequency components associated with anomalous nodes.

\textbf{GHRN}~\cite{gao2023addressing} exploits graph heterophily for anomaly detection by emphasizing high-frequency signals in the graph. It achieves this by selectively pruning edges to highlight class-inconsistent connections, improving the discrimination of anomalous nodes.

\textbf{ARC}~\cite{liu2024arc} is a generalist GAD framework that learns anomaly-aware node representations through an ego-neighbor residual encoder. During inference, ARC employs a cross-attention-based scoring module to estimate node abnormality by leveraging a small set of normal reference nodes.

\textbf{UNPrompt}~\cite{niu2024zero} is a zero-shot GAD method that constructs generalized neighborhood prompts and evaluates node abnormality based on latent attribute predictability, enabling transfer across different graph datasets without task-specific fine-tuning.

\textbf{AnomalyGFM}~\cite{qiao2025anomalygfm} is a generalist graph foundation model designed for anomaly detection. It leverages large-scale pretraining to learn transferable normal and abnormal prototypes and performs anomaly scoring in a zero-shot manner, enabling robust detection across diverse graph datasets.

\begin{table*}[t]
\centering
\caption{Summary of router features used in this work.}
\label{tab:router_features}
\renewcommand{\arraystretch}{1.15}
\begin{tabular}{l l p{10.8cm}}
\toprule
\textbf{Category} & \textbf{Abbr.} & \textbf{Description} \\
\midrule

\multirow{5}{*}{PageRank}
& $\mathrm{PR}_t$
& PageRank score of the target node, reflecting its global importance. \\
& $\mathrm{PR}_{\text{ego\_mean}}$
& Average PageRank value of nodes in the target node's ego graph. \\
& $\mathrm{PR}_{\text{global\_mean}}$
& Mean PageRank value over the entire graph as a global reference. \\
& $\mathrm{PR}_{\text{ego\_rank}}$
& Relative rank of the target node's PageRank within its ego graph. \\
& $\mathrm{PR}_{\text{global\_rank}}$
& Relative rank of the target node's PageRank within the whole graph. \\

\midrule
\multirow{5}{*}{Betweenness}
& $\mathrm{BC}_t$
& Betweenness centrality of the target node, indicating its role as a bridge. \\
& $\mathrm{BC}_{\text{ego\_mean}}$
& Average betweenness centrality of nodes in the ego graph. \\
& $\mathrm{BC}_{\text{global\_mean}}$
& Mean betweenness centrality over all nodes in the graph. \\
& $\mathrm{BC}_{\text{ego\_rank}}$
& Relative rank of the target node's betweenness within its ego graph. \\
& $\mathrm{BC}_{\text{global\_rank}}$
& Relative rank of the target node's betweenness within the whole graph. \\

\midrule
\multirow{5}{*}{Closeness}
& $\mathrm{CC}_t$
& Closeness centrality of the target node, measuring average distance to others. \\
& $\mathrm{CC}_{\text{ego\_mean}}$
& Average closeness centrality of nodes in the ego graph. \\
& $\mathrm{CC}_{\text{global\_mean}}$
& Mean closeness centrality over the entire graph. \\
& $\mathrm{CC}_{\text{ego\_rank}}$
& Relative rank of the target node's closeness within its ego graph. \\
& $\mathrm{CC}_{\text{global\_rank}}$
& Relative rank of the target node's closeness within the whole graph. \\

\midrule
\multirow{6}{*}{Similarity}
& $\mathrm{Sim}_{\text{edge\_avg}}$
& Average cosine similarity between feature representations of connected node pairs. \\
& $\mathrm{Sim}_{1\text{hop}}$
& Mean cosine similarity between the target node and its one-hop neighbors. \\
& $\mathrm{Sim}_{2\text{hop}}$
& Mean cosine similarity between the target node and its two-hop neighbors. \\
& $\mathrm{Sim}_{3\text{hop}}$
& Mean cosine similarity between the target node and its three-hop neighbors. \\
& $\mathrm{Sim}_{4\text{hop}}$
& Mean cosine similarity between the target node and its four-hop neighbors. \\
& $\mathrm{Sim}_{5\text{hop}}$
& Mean cosine similarity between the target node and its five-hop neighbors. \\

\midrule
\multirow{2}{*}{Topology}
& $\mathrm{Deg}_t$
& Degree of the target node, capturing local connectivity. \\
& $\mathrm{Ego}_{\text{size}}$
& Number of nodes in the target node's ego graph. \\

\bottomrule
\end{tabular}
\end{table*}

\section{Implementation Details}
\label{appendix:Implementation}
\subsection{GNN Experts}
\label{appendix:experts_intro}
We employ four expert models in EvoFG, each instantiated with a specific graph propagation mechanism while sharing a unified cross-attention-based integration scheme:

\noindent\textbullet\ \textbf{GCN}~\cite{kipf2016semi} performs neighborhood aggregation based on normalized graph Laplacian to capture local smoothness and homophily in graph-structured data.

\noindent\textbullet\ \textbf{GAT}~\cite{velivckovic2018graph} introduces an attention mechanism into graph message passing, where node representations are updated by aggregating neighbor features with learned, data-dependent attention weights, enabling flexible and anisotropic information aggregation.

\noindent\textbullet\ \textbf{ChebNet} ~\cite{defferrard2016convolutional} approximates spectral graph filters using $K$-order Chebyshev polynomials and captures multi-hop structural information efficiently.

\noindent\textbullet\ \textbf{GPR}~\cite{chien2020adaptive} is built upon generalized PageRank propagation, where node representations are obtained via adaptive combinations of multi-hop message passing with learnable propagation coefficients, allowing flexible control over local  and global information.

Notably, a residual encoder~\cite{qiao2025anomalygfm,liu2024arc} is adopted in GCN, GAT and ChebNet to generate node embeddings. 
To facilitate graph anomaly detection, each GNN expert is followed by an identical cross-attention module~\cite{liu2024arc}, which reconstructs node representations from normal in-context node embeddings, and computes anomaly scores by measuring the distance between the expert output embedding and its reconstructed counterpart.

\subsection{Experimental Hardware}
All models are implemented in Python~3.8 using PyTorch~2.1.2, PyTorch Geometric~2.6.1, and DGL~2.0.0.
Experiments are conducted on a Linux workstation equipped with an Intel Core~i7-12700K CPU and an NVIDIA RTX~A5500 GPU with 24\,GB memory.
CUDA~11.8 is used for GPU acceleration.

\subsection{Hyperparameters}
Unless otherwise specified, all hyper-parameters are fixed across datasets. 
For router optimization, we use a learning rate of $1\times10^{-5}$ and a weight decay of $5\times10^{-5}$. 

The expert models are pretrained separately before router training. Specifically, GCN, GAT, and ChebNet experts are trained for 10 epochs, while the GPR expert is trained for 40 epochs. 
After expert pretraining, the router is warm-up trained for 20 epochs to stabilize routing scores prior to iterative feature generation.

The main training process follows an iterative feature generation–selection–training loop. 

For Shapley-based feature selection, feature contributions are estimated using $T=20$ repeated marginal contribution samples per feature. 
Then, the router is trained for 10 epochs using the selected features. 
For invariant learning, we sample $K=20$ environments and set the variance regularization coefficient to $\lambda=0.8$.

During LLM-driven feature expansion, 15 new router features are generated per iteration, and the entire generation–selection–training loop is repeated for 3 iterations.

\section{Detailed Router Feature Description}
\label{appdenix:router}
\subsection{Initial Router Features}
\label{appendic:router_features}
Table~\ref{tab:router_features} summarizes the set of initial router features used in this work. These features are designed to capture complementary structural and semantic properties of nodes from multiple perspectives, including global and local centrality (PageRank, betweenness, and closeness), neighborhood-aware similarity patterns, and basic topological characteristics. 
For features computed on node ego-graphs, we define the ego-graph of each node as the induced subgraph within its 6-hop neighborhood. 
By combining ego-level statistics, global references, and relative ranking information, the initial feature set provides a diverse yet interpretable foundation for router training before iterative LLM-based feature generation and selection.

\subsection{Operators }



Table~\ref{tab:router_ops} lists the candidate operators used in LLM-based router feature construction. These operators define a flexible yet controlled search space for generating new router features by composing the initial features through unary, binary, and multi-arity transformations. The operator set covers common nonlinear mappings, relative difference and normalization patterns, as well as aggregation statistics, enabling the LLM to express diverse structural and relational patterns while maintaining interpretability and numerical stability.

\begin{table}[t]
\centering
\footnotesize
\caption{Candidate operators used for LLM-based router feature construction. Operators are categorized by arity, and example expressions are illustrated using abbreviated feature names for compactness.}
\label{tab:router_ops}
\renewcommand{\arraystretch}{1.15}
\begin{tabular}{l|l|p{3.5cm}}
\toprule
\textbf{Type} & \textbf{Operator Name} & \textbf{Example (Abbreviated)} \\
\midrule

\multirow{7}{*}{Unary}
& LOG1P
& $\log(1 + \mathrm{BC}_t)$ \\

& LOG
& $\log(\mathrm{Deg}_t)$ \\

& SQRT
& $\sqrt{\mathrm{PR}_t}$ \\

& SQUARE
& $(\mathrm{Sim}_{1\text{hop}})^2$ \\

& CUBE
& $(\mathrm{CC}_t)^3$ \\

& RECIPROCAL
& $1 / \mathrm{Deg}_t$ \\

& SIGMOID
& $\sigma(\mathrm{PR}_t)$ \\

\midrule
\multirow{4}{*}{Binary}
& BINARY\_SUB
& $\mathrm{BC}_t - \mathrm{BC}_{\text{ego\_mean}}$ \\

& BINARY\_DIV
& $\mathrm{BC}_t \,/\, \mathrm{BC}_{\text{ego\_mean}}$ \\

& BINARY\_DIFF\_OVER\_SUM
& $(\mathrm{PR}_t - \mathrm{PR}_{\text{ego\_mean}}) / (\mathrm{PR}_t + \mathrm{PR}_{\text{ego\_mean}})$ \\

& BINARY\_DIV
& $\mathrm{Deg}_t \,/\, \mathrm{Ego\_size}$ \\

\midrule
\multirow{2}{*}{Multi}
& MULTI\_MEAN
& $\mathrm{mean}(\mathrm{Sim}_{1\text{hop}}, \mathrm{Sim}_{2\text{hop}}, \mathrm{Sim}_{3\text{hop}})$ \\

& MULTI\_VAR
& $\mathrm{var}(\mathrm{PR}_t, \mathrm{PR}_{\text{ego\_mean}}, \mathrm{PR}_{\text{global\_mean}})$ \\

\bottomrule
\end{tabular}
\end{table}

\section{Time Complexity of Shapley Estimation}
\label{appendix:shapley_time}
We analyze the computational cost of the proposed Shapley value approximation strategy in Algorithm~\ref{alg:shapley_sampling}.
Let $|\mathcal{F}|=M$ denote the number of router features and let $C_v$ represent the computational cost of evaluating the routing utility function $v(\cdot)$ for a given feature subset.

To perform \emph{exact Shapley computation}, it requires evaluating the marginal contribution of each feature over all possible subsets, leading to a total of $2^{M-1}$ utility evaluations per feature and $\mathcal{O}(M 2^{M})$ overall complexity, which is intractable even for a moderate number of features.

In contrast, our \emph{complementary subset sampling} strategy significantly reduces this cost.
At each iteration $t$, the feature set $\mathcal{F}$ is randomly partitioned into two complementary subsets $S_1$ and $S_2$.
This requires computing $v(S_1)$ and $v(S_2)$ once, followed by $M$ marginal contribution evaluations using reused utilities, as shown in lines 5–8 of Algorithm~\ref{alg:shapley_sampling}.
As a result, the total cost per iteration is $\mathcal{O}(C_v)$ up to a small constant factor.

Over $T$ iterations, the overall complexity becomes $\mathcal{O}(T \cdot C_v)$, which is linear in both the number of sampling iterations and the cost of a single utility evaluation, and notably \emph{independent of the exponential feature subset space}.
In practice, $T \ll 2^{M}$, enabling efficient and scalable feature contribution estimation even when the feature space is continuously expanded by LLM-based generation.

Finally, the post-processing steps for computing the empirical Shapley estimates $\hat{\phi}_i$, standard errors, and $z$-scores incur an additional $\mathcal{O}(TM)$ cost, which is negligible compared to utility evaluation.
Overall, the proposed approximation achieves a favorable trade-off between estimation accuracy and computational efficiency, making it well suited for iterative router feature selection in large-scale settings.
\paragraph{Unbiasedness of the Estimator.}
The Shapley value can be interpreted as the expected marginal contribution of a feature over all possible feature subsets.
In Algorithm~\ref{alg:shapley_sampling}, each random partition induces a uniformly sampled complementary subset for every feature, making the marginal contribution $\Delta_{f_i}^{(t)}$ an unbiased sample of the true Shapley marginal contribution.
Averaging over $T$ independent partitions therefore yields an unbiased estimator of the Shapley value.

\section{Additional Experimental Results}
\label{appendix:e}
\subsection{Performance of Experts}
\label{appendix:experts_experiments}
Table.~\ref{tab:experts} reports the zero-shot GAD performance of individual expert models and the proposed EvoFG across six benchmark datasets. Overall, different experts exhibit distinct strengths depending on dataset characteristics, but none of them consistently performs best across all datasets. Specifically, ARC-based experts achieve strong performance on citation-style datasets such as Cora and CiteSeer, with ChebNet showing particularly high AUROC on Cora, while their performance degrades noticeably on more heterogeneous social networks such as Facebook and Weibo. GPR demonstrates relatively stable performance across datasets and achieves competitive results on BlogCatalog and Weibo, indicating its effectiveness in capturing general structural patterns. However, it still falls short of being universally optimal in the zero-shot setting.

In contrast, EvoFG consistently matches or surpasses the best-performing single expert on most datasets and achieves the strongest overall zero-shot GAD performance. By integrating diverse experts into a Mixture-of-Experts framework and dynamically routing samples to appropriate experts, EvoFG effectively leverages complementary inductive biases from different models. These results demonstrate that combining heterogeneous experts through the proposed routing mechanism leads to more robust and transferable representations, thereby significantly improving zero-shot graph anomaly detection performance compared to any individual expert alone.

\begin{table*}[]
\caption{Expert-wise performance comparison. For each dataset, the best performance per column within each metric is boldfaced, and the second-best is underlined.}
\label{tab:experts}
\begin{tabular}{l|c|cccccc}
\toprule
\multirow{2}{*}{\textbf{Metric}} & \multirow{2}{*}{\textbf{Method}} & \multicolumn{6}{c}{\textbf{Dataset}}                                                                                                                \\
\cmidrule(lr){3-8}

                                 &                                  & Cora                  & CiteSeer              & ACM                   & BlogCatalog           & Facebook                    & Weibo                 \\
\midrule
\multirow{5}{*}{AUROC}           & GAT                     & 80.44 ± 0.62          & 85.50 ± 0.52 & 75.86 ± 0.29 & 73.10 ± 1.25 & \underline{ 64.31 ± 1.18} & 83.36 ± 5.20 \\
                                 & GCN                         & 83.92 ± 0.23          & \underline{ 90.83 ± 0.30}    & 77.09 ± 0.09          & 74.82 ± 0.24          & 63.77 ± 0.24                & 89.75 ± 0.15          \\
                                 & GPR                              & 82.78 ± 0.06          & 89.35 ± 0.19          & 77.58 ± 0.09          & \textbf{75.22 ± 0.09} & 60.60 ± 1.19                & 85.54 ± 0.43          \\
                                 & ChebNet                     & \textbf{ 91.42 ± 0.73}    & 85.80 ± 1.51          & \underline{ 82.65 ± 1.89}    & 72.36 ± 2.30          & 49.06 ± 0.24                & \underline{ 91.42 ± 0.05}    \\
                                 & EvoFG                            & \underline{84.27 ± 0.34} & \textbf{91.57 ± 0.03} & \textbf{91.62 ± 1.42} & \underline{ 75.11 ± 0.47}    & \textbf{66.26 ± 3.24}       & \textbf{93.05 ± 0.88} \\
\midrule
\multirow{5}{*}{AUPRC}           & GAT                     & 37.99 ± 1.39          & 40.81 ± 0.39          & 38.25 ± 0.28          & 34.69 ± 0.80          & 5.58 ± 0.54                 & 49.74 ± 13.06         \\
                                 & GCN                         & 45.42 ± 1.30          & \underline{ 47.79 ± 1.75}    & \underline{ 39.41 ± 0.04}    & \underline{ 36.64 ± 0.23}    & \underline{ 5.94 ± 0.21}           & \underline{ 69.15 ± 0.39}    \\
                                 & GPR                              & 42.33 ± 0.99          & 45.62 ± 0.76          & 39.23 ± 0.16          & \textbf{36.84 ± 0.10} & 5.77 ± 1.24                 & 54.23 ± 1.10          \\
                                 & ChebNet                     & \underline{ 45.43 ± 0.78}    & 26.77 ± 1.49          & 17.76 ± 0.31          & 35.28 ± 0.51          & 2.91 ± 0.05                 & 48.94 ± 0.56          \\
                                 & EvoFG                            & \textbf{47.71 ± 0.99} & \textbf{52.09 ± 1.78} & \textbf{45.78 ± 1.43} & 36.51 ± 0.34          & \textbf{6.09 ± 0.87}        & \textbf{69.80 ± 1.59}  \\
\bottomrule
\end{tabular}
\end{table*}

\subsection{Further Hyper-Parameter Analysis}

\begin{figure}[t]
    \centering
    \includegraphics[width=0.75\columnwidth]{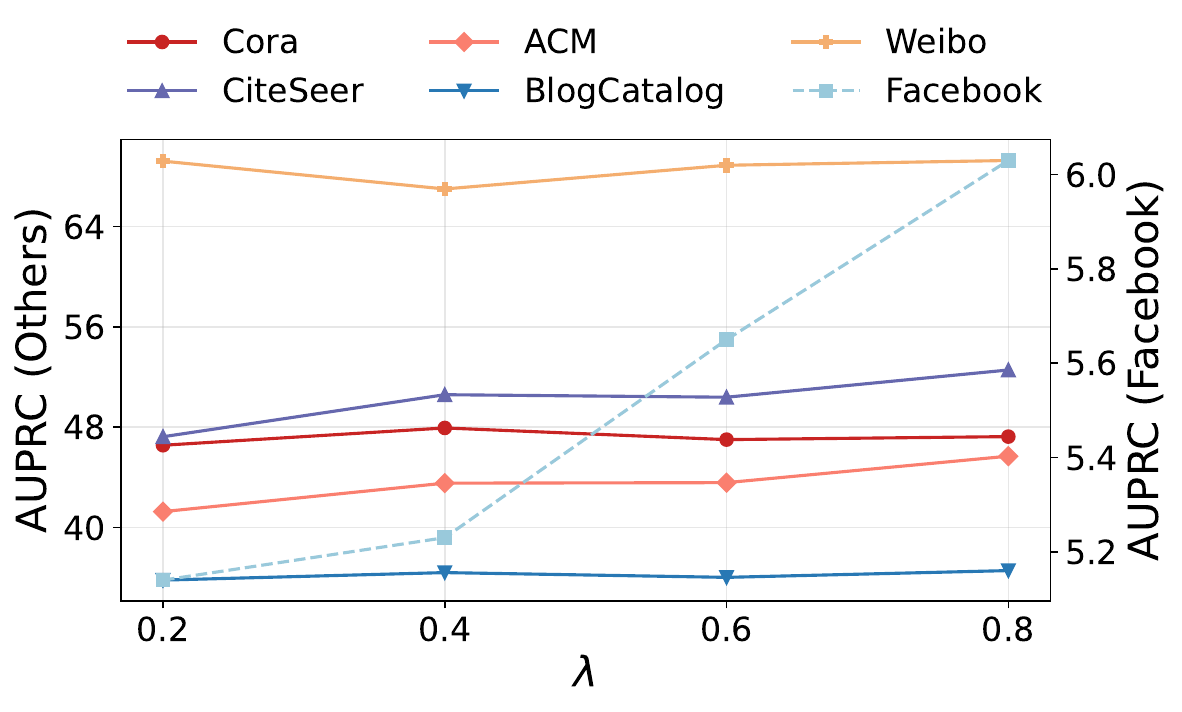}
    \caption{Coefficient $\lambda$ of invariant learning in \textbf{EvoFG (Full)} across different datasets.}
    \label{fig:invar_lamda}
\end{figure}

We further investigate the sensitivity of our method to two key hyper-parameters in the invariant learning component: the invariance regularization coefficient $\lambda$ and the number of sampled environments $K_{\text{env}}$.
Figure~\ref{fig:invar_lamda} analyzes the effect of $\lambda$, which controls the trade-off between empirical risk minimization and invariance regularization across environments.
As $\lambda$ increases from small values, the performance on most datasets consistently improves, indicating that enforcing invariance helps the router suppress environment-specific spurious correlations and learn more stable routing patterns.
However, overly large values lead to marginal gains or saturation, suggesting that excessive invariance constraints may restrict the router’s flexibility.
In practice, a moderate coefficient (e.g., $\lambda=0.6\sim0.8$) achieves a favorable balance between robustness and expressiveness.

\begin{figure}[t]
    \centering
    \includegraphics[width=0.75\columnwidth]{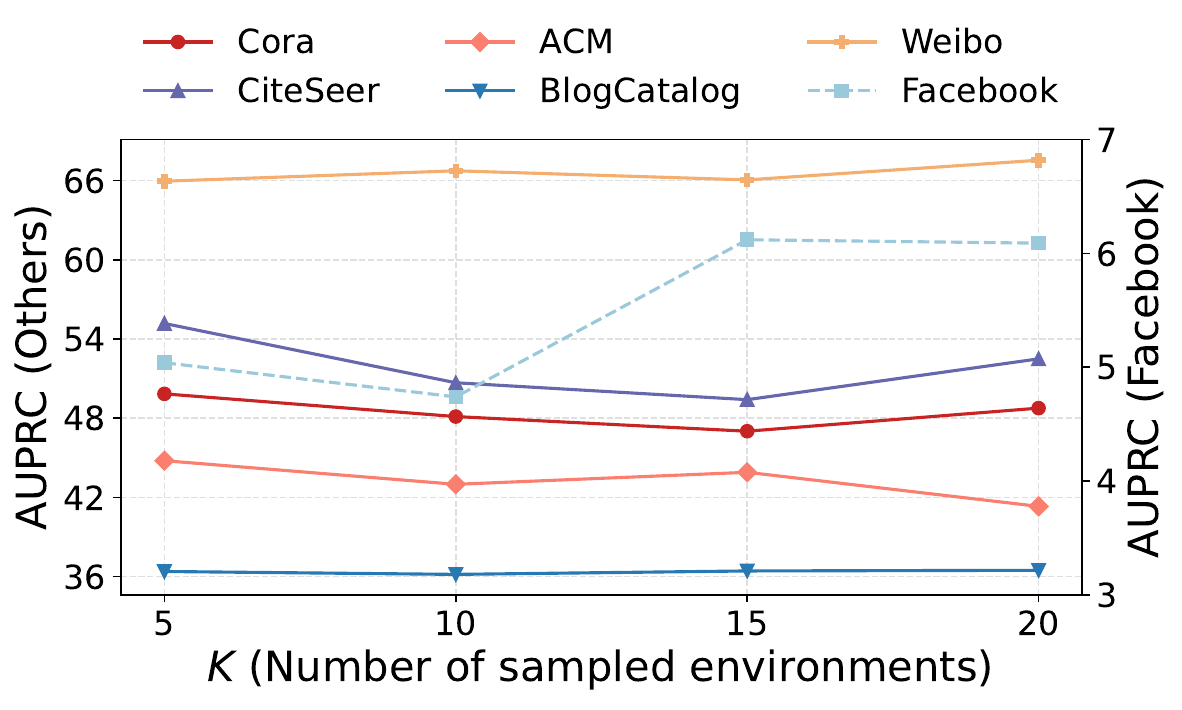}
    \caption{Number of environments of invariant learning in \textbf{EvoFG (Full)} across different datasets.}
    \label{fig:invar_K_env}
\end{figure}

Figure~\ref{fig:invar_K_env} studies the impact of the number of simulated environments $K_{\text{env}}$.
Increasing $K_{\text{env}}$ generally improves performance by exposing the router to more diverse environment perturbations, thereby enhancing its ability to capture invariant routing signals.
Nevertheless, the performance gain gradually diminishes as $K_{\text{env}}$ becomes large and tends to stabilize beyond a certain point (e.g., $K_{\text{env}}\geq15$).
This observation suggests that a sufficient but not excessive number of environments is enough to model dominant environmental variations while avoiding unnecessary computational overhead.
Based on these results, we set $K_{\text{env}}=20$ in all experiments.

\subsection{Further Case Studies}
\label{appendix:selection}

\begin{figure*}[t]
    \centering
    \begin{subfigure}[t]{0.23\linewidth}
        \centering
        \includegraphics[width=\linewidth]{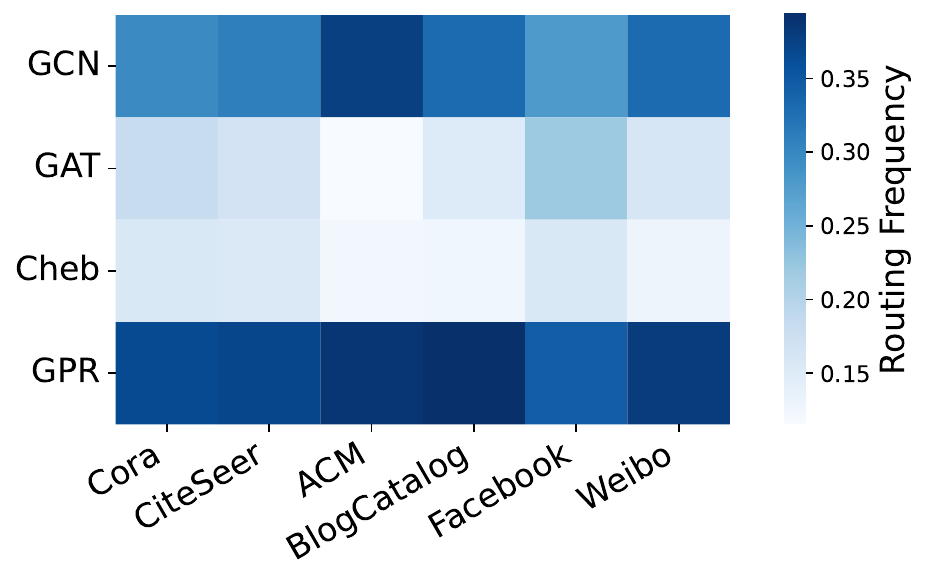}
        \caption{FeatEvo $\rightarrow$ RandFeat}
    \end{subfigure}
    \hfill
    \begin{subfigure}[t]{0.23\linewidth}
        \centering
        \includegraphics[width=\linewidth]{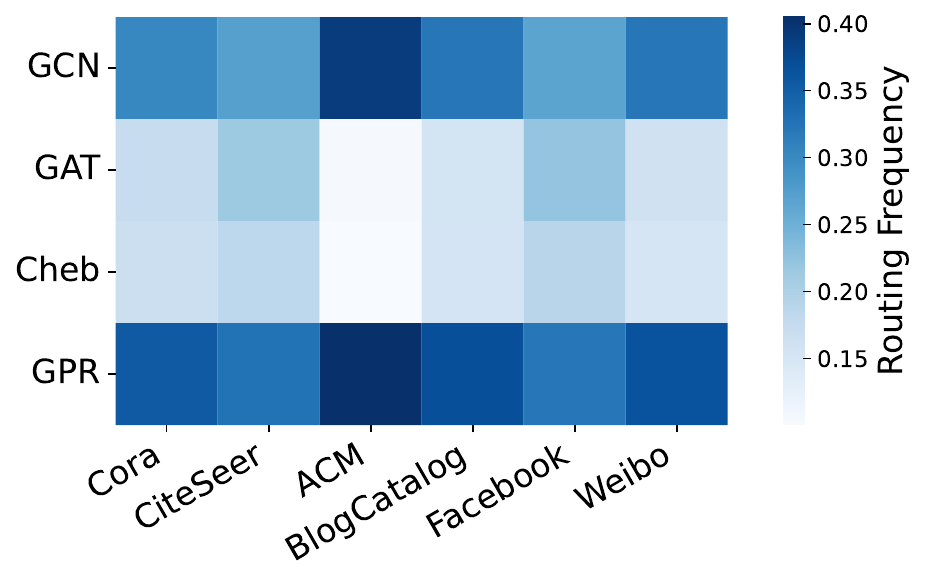}
        \caption{w/o Feature Selection}
    \end{subfigure}
    \hfill
    \begin{subfigure}[t]{0.23\linewidth}
        \centering
        \includegraphics[width=\linewidth]{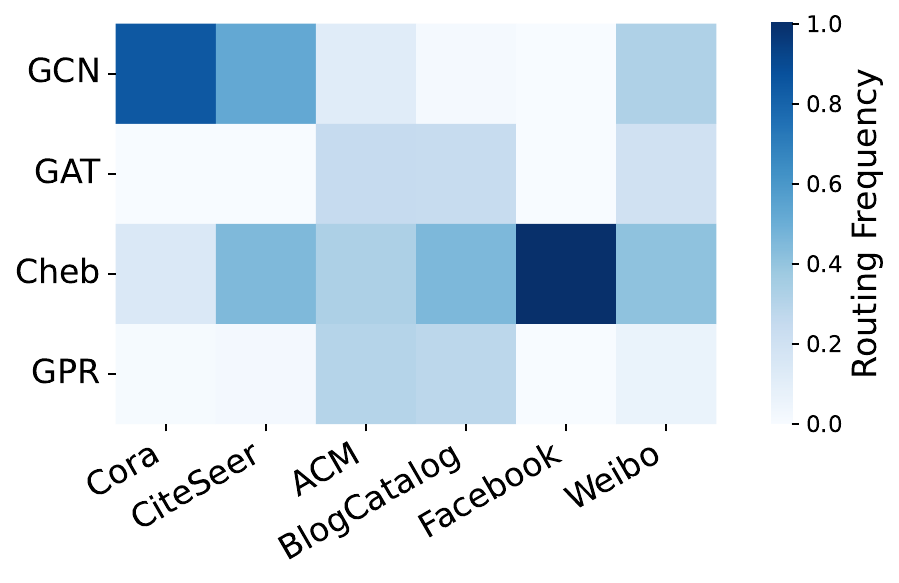}
        \caption{w/o Router Memory}
    \end{subfigure}
    \hfill
    \begin{subfigure}[t]{0.23\linewidth}
        \centering
        \includegraphics[width=\linewidth]{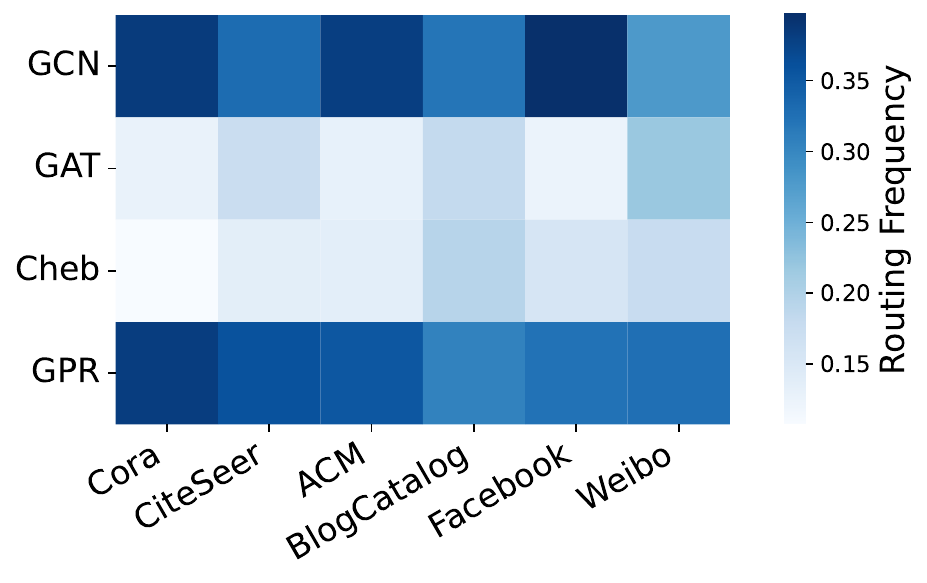}
        \caption{w/o Invariant Learning}
    \end{subfigure}

    \caption{Expert soft routing frequency heatmaps for EvoFG's ablation variants.}
    \label{fig:ablation_heatmap}
\end{figure*}

\noindent\textbf{Soft Routing Frequency.}
To characterize the routing behavior of the MoE router, we define the \emph{soft routing frequency}, which jointly accounts for both the routing scores assigned to experts and how frequently experts are selected across samples.

Given a test dataset with $N$ samples and $E$ experts, let $\mathbf{P}[i,e]$ denote the routing weight assigned to expert $e$ for sample $i$. We first accumulate the soft routing mass for each expert:
\begin{equation}
M_e = \sum_{i=1}^{N} \mathbf{P}_{i,e}.
\end{equation}
The soft routing frequency is then defined by normalizing the accumulated mass:
\begin{equation}
f_e^{\mathrm{soft}} =
\frac{M_e}{\sum_{e'=1}^{E} M_{e'}}.
\end{equation}
Under standard soft routing where $\sum_{e=1}^{E} \mathbf{P}_{i,e}= 1$, the denominator equals $N$, and $f_e^{\mathrm{soft}}$ corresponds to the average routing weight assigned to expert $e$ over the dataset.

\noindent\textbf{Additional Analysis of Routing Behaviors.}
In addition to the routing analysis of the full model presented in Sec.~\ref{case1}, we further examine the routing behaviors of several ablation variants using soft routing frequency.
This supplementary analysis provides a more comprehensive understanding of how different components of EvoFG influence the stability and diversity of expert selection.

\textbf{w RandFeat} shows a much less discriminative routing pattern. The router predominantly favors GPR and ARC, while the other two experts are consistently under-selected. Moreover, the soft routing distributions become highly similar across datasets, suggesting that random feature selection fails to provide informative signals for distinguishing dataset-specific characteristics. As a result, the router cannot effectively personalize expert selection for different data distributions.

The routing behavior of \textbf{w/o Feature Selection} is the closest to that of the full model. The router still assigns varying expert proportions across datasets, reflecting a certain degree of distribution-aware routing. However, subtle differences remain compared to EvoFG (Full), and these deviations lead to slightly inferior performance. This observation confirms that feature selection, although not solely decisive, helps refine the router feature space and enables more accurate expert selection.

For \textbf{w/o Router Memory}, the routing behavior collapses severely. On several datasets, some experts are selected with near-zero probability, and on the Facebook dataset, the router almost exclusively selects ARC$\_$Cheb for nearly all samples. Such extreme and unstable routing patterns indicate that without memory, the router fails to accumulate and reuse routing knowledge, resulting in a breakdown of meaningful expert selection.

Finally, \textbf{w/o Invariant Learning} exhibits a clear expert collapse phenomenon. The router consistently over-selects ARC and GPR while largely ignoring the other experts across datasets. This behavior suggests that without invariant learning, the router overfits to spurious correlations and fails to learn invariant routing principles that generalize across data distributions, preventing effective dataset-specific expert personalization.

Overall, this case study demonstrates that each component of EvoFG contributes to learning stable, diverse, and distribution-aware routing behaviors, and their combination is crucial for robust zero-shot expert selection.

\section{Related Work}
\label{RW}
\subsection{Generalist GAD}
Due to the limited availability of anomaly annotations in many graph datasets, generalist graph anomaly detection has attracted increasing attention in recent years, leading to the emergence of several representative approaches.
ARC~\cite{liu2024arc} is the first method proposed for generalist GAD prediction, with a focus on few-shot prediction. ARC proposed a cross-attention mechanism for reconstructing node embedding from normal in-context node embeddings, and then computes anomaly scores by measuring the distance between the original node embedding and the reconstructed counterpart.
UNPrompt~\cite{niu2024zero} firstly pre-trains a GNN model in a contrastive learning manner, and then a group of learnable generalized neighborhood prompts is aggregated to the node features, and they are together fed into the pre-trained GNN model for zero-shot GAD prediction.
AnomalyGFM~\cite{qiao2025anomalygfm} pre-trains a GNN model to learn data-independent normal and abnormal prototypes, then the prototypes can be used in few/zero-shot GAD prediction. 
Different from the above methods, which only use a single GNN model to extract node embeddings, our proposed EvoFG can adaptively select appropriate GNN experts across different graph domains in zero-shot inference. 

\subsection{MoE in Graph Mining}
Mixture-of-Experts (MoE) is a classical paradigm for conditional computation, where a gating network dynamically routes inputs to a subset of specialized experts, enabling scalable model capacity and expert specialization~\cite{shazeer2017outrageously}.

Building upon this paradigm, MoE mechanisms have been increasingly adopted in graph mining tasks to address structural heterogeneity and diverse learning objectives. In link prediction, MoE-based graph models dynamically select experts to capture different relational patterns and improve predictive performance~\cite{ma2024mixture}. For graph classification, MoE frameworks allocate graphs or substructures to specialized GNN experts, allowing the model to adapt to varying graph-level characteristics~\cite{hu2021graph}. To improve scalability on large-scale graphs, MoE has been employed to adaptively select aggregation scopes, such as first-order or second-order neighborhoods, thereby reducing redundant computation while preserving expressive power~\cite{wang2023graph}.

MoE has also shown strong effectiveness in spatio-temporal and traffic-related graph learning tasks. In traffic forecasting and imputation, MoE-based models dynamically route spatio-temporal patterns to specialized experts to handle non-stationarity and evolving traffic dynamics~\cite{lee2024continual,he2024st,wang2025stamimputer}. For general spatio-temporal graph modeling, MoE designs have been proposed to incorporate tailored GNN filters that better capture spatial dependencies and temporal dynamics~\cite{lee2024testam}. Beyond forecasting tasks, MoE has been applied to time-series anomaly detection, where memory-augmented routers are introduced to enhance long-term temporal dependency modeling and improve anomaly sensitivity~\cite{huang2025graph}.

Despite these advances, existing graph MoE studies primarily focus on prediction or forecasting tasks and rarely investigate MoE architectures for graph anomaly detection (GAD), especially under zero-shot or cross-dataset settings. This motivates our work on MoE-based GAD, where expert specialization and adaptive routing are explicitly leveraged to improve generalization across diverse graph domains.

\subsection{Feature Engineering with LLM}

Classical feature engineering methods construct new features through predefined feature transformations, which often follow fixed and heuristic paradigms.
Such approaches can significantly limit the exploration of the feature space and restrict the expressiveness of the resulting representations~\cite{horn2019autofeat, kanter2015deep,khurana2018feature,wang2025id}.
Recently, large language models (LLMs) have demonstrated strong capabilities in automated feature engineering by leveraging the extensive domain knowledge learned during pre-training.

Several recent studies have explored LLM-based feature engineering.
LFG~\cite{zhang2024dynamic} proposes a dynamic and adaptive feature generation framework that combines LLM-driven feature and operator selection with Monte Carlo Tree Search to identify effective new features.
LLM-FE~\cite{abhyankar2025llm} and OCTree~\cite{nam2024optimized} iteratively generate feature transformation hypotheses, using downstream task performance as a reward signal to guide feature evolution.
FeRG-LLM~\cite{ko2025ferg} further enhances feature generation by constructing two-stage conversational datasets and fine-tuning LLMs with direct preference optimization (DPO), enabling the model to provide rationale-aware feedback when generating feature transformation ideas and executable code.

Despite their effectiveness, existing LLM-based feature engineering methods are primarily designed for tabular data, where each feature column is associated with explicit semantic descriptions.
Such structured semantics allow LLMs to directly leverage their domain knowledge to propose meaningful feature transformations.
In contrast, graph anomaly detection datasets typically consist of node features that encode heterogeneous information from diverse domains, often without clear semantic annotations.
As a result, directly applying existing LLM-based feature engineering techniques to generalist GAD is non-trivial and may lead to unreliable or misleading features, making these methods ill-suited for generalist graph anomaly detection.

\end{document}